\documentclass[aps,prd,reprint,groupedaddress,nofootinbib]{revtex4-1}

\usepackage{graphicx}
\usepackage{multirow} 
\usepackage{amssymb}
\usepackage{amsmath}
\usepackage{enumerate}
\usepackage{todonotes}
\usepackage{hyperref}

\begin{document}


\title{A Comprehensive Approach to Tau-Lepton Production by High-Energy Tau Neutrinos 
Propagating Through Earth} 
\author{Jaime Alvarez-Mu\~niz$^1$, Washington R. Carvalho Jr.$^{2,1}$, K\'evin Payet$^3$, Andr\'es Romero-Wolf$^4$, Harm Schoorlemmer$^5$, Enrique Zas$^1$}
\affiliation{$^1$Departamento de F\'\i sica de Part\'\i culas \& Instituto Galego de F\'\i sica de Altas Enerx\'\i as, Univ. de Santiago de Compostela, 15782 Santiago de Compostela, Spain\\
$^2$Departamento de F\'\i sica, Universidade de S\~ao Paulo, S\~ao Paulo, Brazil\\
$^3$ Universit\'e Joseph Fourier (Grenoble I); Currently at La Javaness, 75010, Paris, France\\
$^4$ Jet Propulsion Laboratory, California Institute of Technology, Pasadena, CA 91109, USA\\
$^5$ Max-Planck-Institut f\"ur Kernphysik, 69117, Heidelberg, Germany}





\begin{abstract}
There has been a recent surge in interest in the detection of $\tau$ lepton-induced air showers from detectors at altitude. When a $\tau$ neutrino ($\nu_\tau$) enters the Earth it produces $\tau$ leptons as a result of nuclear charged current interactions. In some cases, this process results in a $\tau$ lepton exiting the surface of the Earth, which can subsequently decay in the atmosphere and produce an extensive air shower. These upward-going air showers can be detected via fluorescence, optical Cherenkov, or geomagnetic radio emission. Several experiments have been proposed to detect these signals. We present a comprehensive simulation of the production of $\tau$ leptons by $\nu_\tau$'s propagating through Earth to aid the design of future experiments. These simulations for $\nu_\tau$'s and leptons in the energy range from $10^{15}$ eV to $10^{21}$ eV treat the full range of incidence angles from Earth-skimming to diametrically-traversing. Propagation of $\nu_\tau$'s and leptons include the effects of rock and an ocean or ice layer of various thicknesses. The interaction models include $\nu_\tau$ regeneration and account for uncertainties in the Standard Model neutrino cross-section and in the photo-nuclear contribution to the $\tau$ energy loss rate. 
\end{abstract}

\keywords{Tau Neutrinos; Air Shower}
\maketitle


\section{Introduction}

Despite decades of observation, including currently operating state-of-the-art detectors~\cite{Auger_Inst_2015, TA_Inst_2012}, the origin of ultra-high energy (UHE, $\gtrsim$~$10^{18}$~eV) cosmic rays remains unknown with source model parameters relatively unconstrained~\cite{Auger_MCMC_2017}. 
The propagation of UHE cosmic rays through the cosmic photon backgrounds is expected to produce a flux of UHE neutrinos~\cite{Greisen_1966, Zatsepin_Kuzmin_1966, Berezinsky_Zatsepin_1966, Kotera_2010}. 
The detection of or tighter limits on this cosmogenic UHE neutrino flux has the potential to significantly constrain models of UHE cosmic-ray sources~\cite{Seckel_2005,Auger_Neutrino_2015,IceCube_UHE_2016}. 

UHE neutrinos have been searched for in the past decades with several experiments using the Moon~\cite{Hankins_1996, Gorham_2004, Bray_2015}, the Earth's crust with the Pierre Auger Observatory~\cite{Auger_Neutrino_2015}, and the Antarctic ice cap with IceCube~\cite{IceCube_UHE_2016} and ANITA~\cite{Gorham_nu_2010}, each producing the most constraining limits to date in different parts of the UHE band. 
Recently an extra-terrestrial flux of neutrinos extending above $10^{15}$~eV has been detected with IceCube~\cite{IceCube_detection_2014}. 
The origin of this flux is unknown and while it is dominated by extragalactic sources~\cite{Denton_2017} it is not expected to be due to the UHE cosmic ray interaction with cosmic photon backgrounds~\cite{Roulet_2013, Anchordoqui_2014}. It is also unknown the extent to which the detected flux can be extrapolated to the UHE band or if there is a cutoff energy \cite{Anchordoqui_2017}. 

Although cosmogenic neutrinos are expected to be produced predominantly in electron and muon flavors, flavor mixing over cosmic propagation distances predicts a significant fraction ($\sim1/3$) of $\tau$ neutrinos~\cite{Athar_2000, Learned_Pakvasa_1995}. The $\tau$ neutrinos ($\nu_\tau$) provide a detection channel that is unlike the electron and muon counterparts. When a $\nu_\tau$ interacts via a charged current interaction it produces a $\tau$ lepton along with a hadronic shower. 
The $\tau$ lepton has an extremely short lifetime compared to the muon, compensating the enormous Lorentz boost at UHE, so that the $\tau$ is likely to decay before losing a significant amount of its energy. One such consequence are the so-called double bang events~\cite{Learned_Pakvasa_1995} expected by experiments searching for showers in ice such as IceCube. 
Another promising channel for detection, where $\tau$ leptons produced in the Earth exit into and decay in the atmosphere producing an up-going extensive air shower,
was originally studied in~\cite{Fargion_1999, Letessier-Selvon_2001}. These up-going air showers could be potentially observed by the surface detector of the Pierre Auger Observatory \cite{Auger_Neutrino_2015} as an Earth-skimming shower~\cite{Bertou_2002, Feng_2002} and potentially observable from a high-altitude observatory such as ANITA~\cite{Gorham_2016}. Recently, several experiments are being proposed to search for this signal such as Ashra \cite{ASHRA_2013} and CHANT~\cite{Neronov_2017} using optical Cherenkov 
detection, and GRAND~\cite{Martineau-Huynh_2017} using a ground array of radio detectors. 

The propagation of $\nu_\tau$'s entering the Earth resulting in a $\tau$ lepton exiting into the atmosphere is a complex problem involving multiple scales. It involves the combined effect of the $\tau$ lepton lifetime, its energy loss, and the neutrino cross-section, each having a different dependence on energy together with the available matter depth which is a rapidly varying function of the $\nu_\tau$ incidence angle (see for instance~\cite{Zas_2005} for a review). 
As the neutrino propagates through the Earth the $\nu_\tau$ flux is attenuated because of interactions. However, there is also a ``regeneration" effect~\cite{Saltzberg_1998, Blanch_Bigas_2008}: Charged-current interactions followed by $\tau$-lepton decays (in which a lower energy $\nu_\tau$ is always produced) as well as neutral current interactions 
shift down the neutrino energy ``regenerating" the $\nu_\tau$ flux.
The effect of regeneration depends heavily on the $\nu_\tau$ incidence angle. It can be negligible for Earth-skimming trajectories but result in well over a factor of two shift in $\nu_{\tau}$ energy for incidence angles with higher depth.
In addition, it has been shown that because of competition between energy loss and decay, the density of the matter just before the $\tau$ lepton exits the Earth can have a significant impact on the rate of exiting $\tau$ leptons which is important for observation near the ocean or the ice caps~\cite{Palomares-Ruiz_2006}.

The $\tau$ production mechanism also depends crucially on uncertainties on the 
neutrino-nucleon cross-section as well as the $\tau$ lepton energy loss rate
at UHE, both of which have to be extrapolated to kinematical regions 
which are well outside those that have been explored in accelerators.
Several publications have studied the propagation and detectability of $\tau$ leptons exiting the Earth into the atmosphere 
in different energy and angular ranges, using Monte Carlo simulations and/or semi-analytical techniques and 
with a focus on the observability in underwater/ice and/or ground-level arrays of particle and/or fluorescence detectors
\cite{Palomares-Ruiz_2006, Bottai_2001, Fargion_2002, Dutta_2002, Tseng_2003, Bottai_2003, Montaruli_2004, Yoshida_2004, Aramo_2005, Dutta_2005, Gora_2007, Blanch_Bigas_2008, Jeong_2017}. 

The objective of the comprehensive simulations presented in this paper is to provide useful results  for the design of future detectors including the effects that are relevant to observations over rock, ocean, or ice from the ground or at altitude, while considering uncertainties in the interaction models. We present the characteristics and results obtained with a simulation based on \cite{Payet_PhD} for estimating the probability and energy distribution of $\tau$ leptons exiting the surface of the Earth given a UHE $\nu_\tau$ flux entering at a given incidence angle. The simulation accounts for the particle interactions, energy loss and decay processes relevant to the propagation of $\nu_\tau$'s through the Earth, including $\nu_\tau$ regeneration, as well as for the geometry of the propagation including the features of the  Earth density profile (with spherical symmetry), and settable ocean/ice layer thickness all in a single simulation package. 

The simulation can be used in a wide energy and angular range relevant for most experiments, namely, from $10^{15}$ eV to $10^{21}$ eV in $\nu_\tau$ energy and the full incidence angle range $0^\circ-90^\circ$. A recent study~\cite{Jeong_2017} produced simulations with detailed models of neutrino interaction cross-sections and $\tau$-lepton energy loss models with consistent account of inelasticity. The simulations presented here also allow for variations of the cross-section and energy loss model, although in not as detailed a manner (see Sec. III). However, we also include the effects of a layer of ice or water as well as regeneration, which were not included in~\cite{Jeong_2017}. While these effects may not be relevant to estimating the sensitivity of the Pierre Auger Observatory (as done in~\cite{Jeong_2017}), they may be relevant to interpreting $\tau$-lepton air-showers observed at altitude, as mentioned in~\cite{Gorham_2016}. 

The simulation package used to produce the results of this paper is being made publicly available\footnote{\url{https://github.com/harmscho/NuTauSim}}. The simulation allows the user to specify functions for neutrino-nucleon and photo-nuclear contribution to tau energy loss curves that account for Standard Model uncertainties. The executable takes in the incident $\nu_\tau$ energy, direction, number of injected neutrinos, cross-section and energy loss model in the command line, and produces a list of $\tau$-lepton energies that exit into the atmosphere. A novel feature of the simulation is that it tracks the chain of $\nu_\tau$ interactions and  $\tau$ lepton decays that lead to the production of an exiting $\tau$ lepton, shedding light on their production mechanism as a function of incidence angle and the initial $\nu_\tau$ energy. The simulation has been designed with the intention that other models, more detailed accounting of particle interactions, or additional media effects may be easily included for future developments. 

%

The paper is organized as follows: 
in Section II we provide a summary of $\nu_\tau$ propagation through the Earth. 
In Section III we review some of the Standard Model uncertainties in the neutrino-nucleon cross-section and $\tau$ lepton energy-loss rate. 
In Section IV we give results for the dependence of the exiting $\tau$ lepton flux on $\nu_\tau$ energy,  $\nu_\tau$ regeneration, 
ice layer thickness, and particle interaction models. 
In Section V we provide a discussion of the results and conclusions. 

\section{Tau Neutrino Propagation Through the Earth}
\label{sec:propagation}
\begin{figure}[t]
  \centering
   \includegraphics[width=1.0\linewidth]{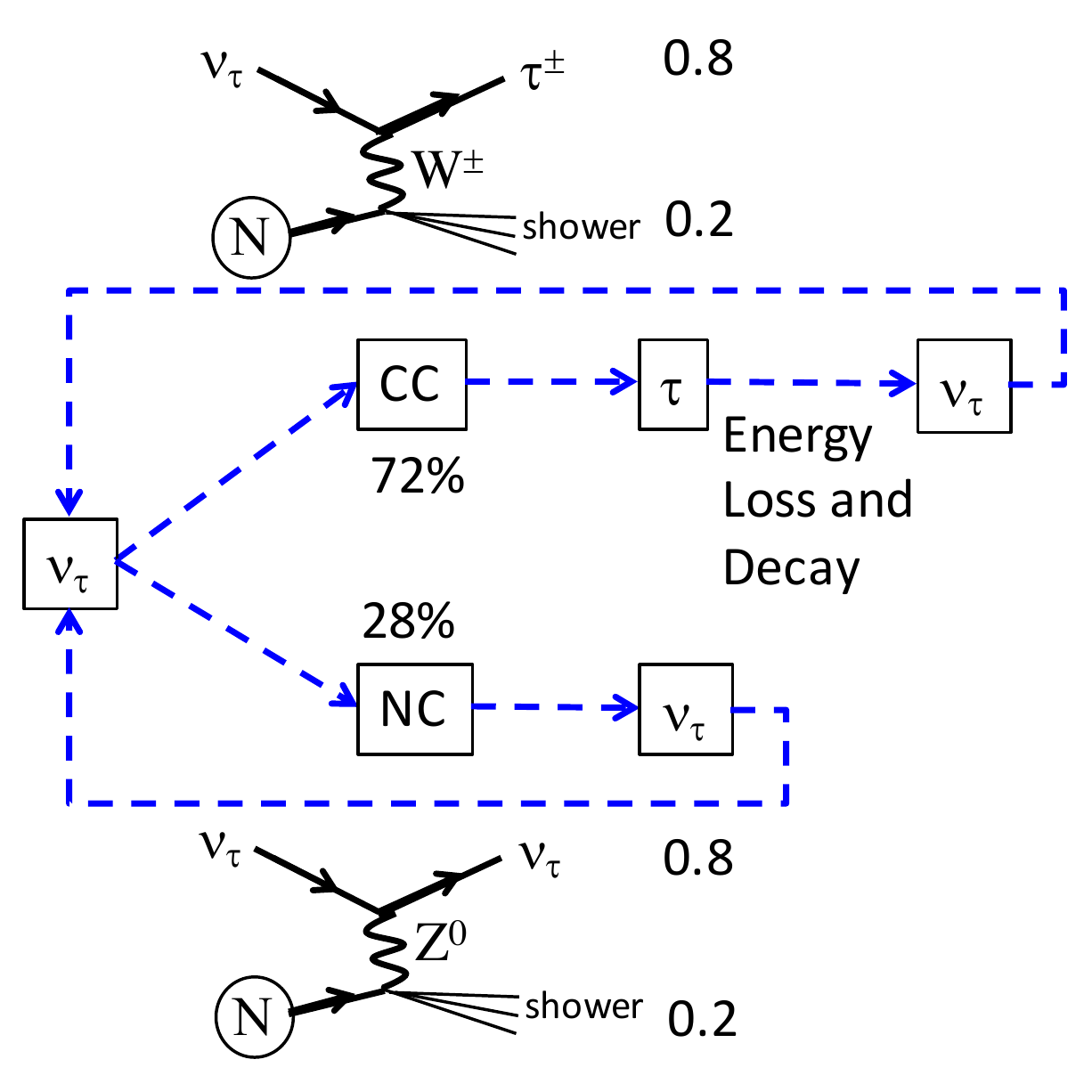} %
   \caption{Diagram of the $\nu_\tau$ propagation process. A $\nu_\tau$ can interact either via charged current (CC) or neutral current (NC) interaction. The CC interaction occurs $\sim$72\% of the time and results in the production of a $\tau$ lepton, carrying on average $\simeq 80\%$ of the energy of the $\nu_\tau$ at UHE and loses energy while propagates. The $\tau$ lepton can escape the Earth before it decays. If it decays in the Earth, it will produce a $\nu_\tau$, which will continue to propagate. NC interactions occur $\sim$28\% of the time and produce a $\nu_\tau$ which also continues to propagate. See text for more details. }
   \label{fig:tau_prop}
\end{figure}

As a $\nu_\tau$ of energy $E_\nu$ propagates through the Earth it will primarily interact with nucleons, which can result in the production of a $\tau$ lepton. 
The processes involved in $\nu_\tau$ and $\tau$ lepton propagation are shown in Figure~\ref{fig:tau_prop}. A neutral 
current (NC) interaction occurs $\sim 28 \%$ 
of the time resulting in the production of a hadronic shower and a $\nu_\tau$. 
The secondary $\nu_\tau$ carries a fraction of the energy given by
$(1-y)E_\nu$. The lab-frame inelasticity $y$ is a standard dimensionless variable corresponding to the fraction of the neutrino energy 
carried away by the hadronic shower. The $y$-distribution is non-centralized with an average $\langle y \rangle \simeq 0.2$ at UHE \cite{Gandhi_1998} and large variance. The hadronic shower energy is absorbed in a few tens of meters in the Earth while the $\nu_\tau$ continues to 
propagate. A charged current interaction (CC) occurs $\sim 72 \%$ 
of the time producing a hadronic shower and a $\tau$ lepton. 
The average fraction of the energy carried away by the hadronic shower at UHE is also $\simeq 0.2 \ E_\nu$ with a similar 
$y$ distribution.

As the produced $\tau$ lepton propagates through the Earth it will lose energy 
via ionization, pair-production, and photo-nuclear interactions. 
The $\tau$ lepton will decay with a lifetime of 0.29~ps in its 
rest frame, which, for a given energy $E_\tau$, corresponds to a 
{\it decay-length} of $(E_\tau/10^{17}\ \mbox{eV})\times4.9$~km. 
At low energies the $\tau$ lepton tends to decay before losing a significant 
amount of energy and the $\tau$ range increases linearly with energy because 
of the Lorentz boost. 
However, this behavior changes above an energy scale typically in the range
between $10^{17}$~eV and $10^{18}$~eV~\cite{Zas_2005}. In this regime energy losses decrease 
the $\tau$-lepton energy until it either decays or exits the Earth, depending 
on the available matter depth.
If the $\tau$ lepton decays inside the Earth, it produces a variety of 
particles, depending on the decay mode, always including a regenerated 
$\nu_\tau$ with lower energy that continues to propagate. 


When the $\tau$ lepton exits the Earth's crust into the 
$\sim1000$ times lower density atmosphere, it undergoes negligible energy 
loss. The dominating process in this case is  $\tau$-decay, 
described by an exponentially falling distribution characterized by 
the decay-length.
The decay products result in hadrons $>60\%$ of the time or 
in an electron with a  branching ratio of $17.9\%$. In both
cases an extensive air shower is produced provided the decay happens in the 
atmosphere.
The decay can also result in a muon $17.9\%$ of the time, which is unlikely 
to result in an extensive air shower. 
The $\tau$ lepton decays in the atmosphere and their 
subsequent air showers are not handled by the Monte 
Carlo simulations presented here. 

\begin{figure}[t]
  \centering
   \includegraphics[width=0.7\linewidth]{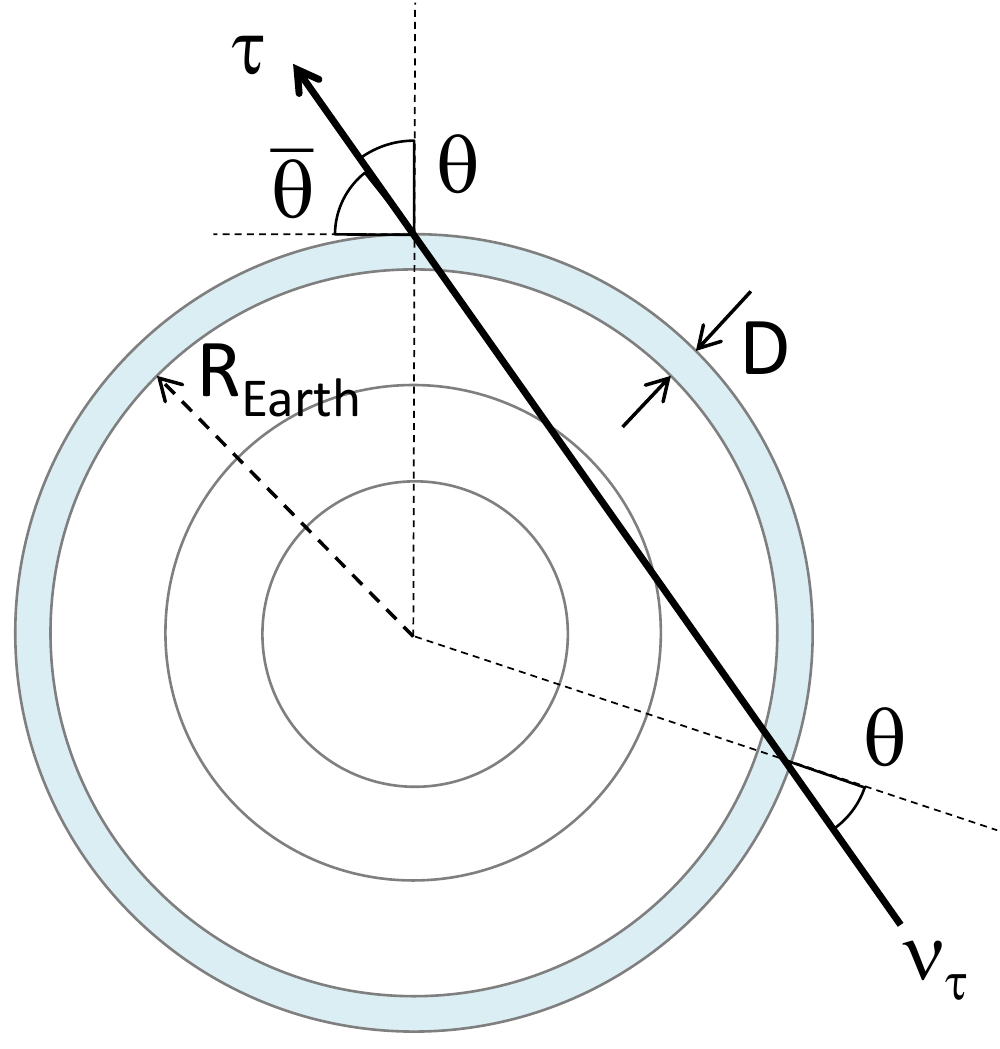} %
   \caption{Simulation geometry for a $\nu_\tau$ entering the Earth with
     incidence angle $\theta$. The model of the Earth assumes a sphere of
     radius $R_{\rm Earth}$ and settable ocean or ice depth $D$. The
     $\nu_\tau$ is injected with a specified exit angle $\theta$, which
     determines the trajectory it will take across the various subsurface
     layers of varying density where the interactions will take place. Results
     are given in terms of the emergence angle $\bar \theta= 90^\circ -
     \theta$ which is the complement of the exit and entry angles.}
   \label{fig:tau_geom}
\end{figure}

The relevant geometric parameters of the simulation are shown in the diagram in Figure~\ref{fig:tau_geom}. The direction of the injected neutrino is determined by the incidence angle $\theta$. For our spherically symmetric model, the exit angle is identical to the incidence angle. The complement is the emergence angle $\bar{\theta}=90^{\circ}-\theta$. The Earth is modeled as a sphere of radius $R_{\rm Earth}$ with subsurface layers of varying density according to the Preliminary Earth Reference Model~\cite{PREM_model}. An ocean or ice layer of thickness $D$ can be specified in the simulation. 

The interactions of the injected particle are randomly sampled while it 
propagates. The depth traversed by the $\nu_\tau$ before interacting is 
obtained from an exponential distribution with mean free path determined by 
the total cross-section as a function of $E_\nu$. Whether the interaction is 
CC or NC is randomly chosen according to their relative weight on the total 
cross-section (${\rm CC/NC} \simeq 2.6/1$ at UHE). In case of 
a NC interaction, the $\nu_\tau$ propagation proceeds with the new value of 
$E_\nu$ sampled from the $y$-distribution. In case of a CC interaction, the 
energy of the resulting $\tau$ lepton is also sampled from the 
 same
$y$-distribution. The $\tau$ lepton propagation accounts for energy loss in 
small depth steps and at each step the probability of decay is evaluated at 
the new $\tau$ lepton energy and randomly sampled from an exponential 
distribution. 
If the $\tau$ lepton decays the energy of the resulting $\nu_\tau$ is sampled from the corresponding distribution obtained with the TAUOLA package \cite{TAUOLA} with an average value of $\langle E_\nu\rangle \simeq 0.3 E_\tau$. The simulation stops when either the propagating $\nu_\tau$ or $\tau$ lepton exits the Earth into the atmosphere. In the case of an exiting $\tau$ lepton, the simulation records its final energy, emergence angle, the initial $\nu_\tau$ energy, the number of CC interactions (at least one), NC interactions, and $\tau$ lepton decays (if any). This feature of the simulation allows us to study in detail the 
chain of events that typically lead to the production of $\tau$ leptons in the atmosphere as a function of the initial $\nu_\tau$ energy $E_\nu$ and emergence angle $\bar\theta$.

\section{Neutrino Interaction and Tau Energy-Loss Models}
\label{sec:uncertainties}

The main source of uncertainty in the calculation of the $\tau$-lepton fluxes arises because the neutrino cross section and the $\tau$ energy loss are needed in an energy range where there is no experimental data. 
Neutrino cross-sections and $\tau$ energy losses are calculated using the Standard Model and require extrapolating the structure functions of the nucleons to kinematical regions that are unexplored. Structure functions depend on the four-momentum of the nucleon $p$ and the four-momentum transferred to the nucleon $q$ but they are typically described in
terms of the two independent variables $Q^2=-q^2$ and $x=Q^2/(2 p \cdot q)$. The kinematical region mostly contributing to the neutrino cross-section and $\tau$ lepton energy losses are different. While in both regions the relevant values of $x$ are extremely low, for the neutrino cross-section $Q^2 \sim M_W^2 \sim 6400~{\rm GeV}^2$, of order the mass squared of the exchanged weak bosons, while for the $\tau$ lepton energy loss $Q^2$ is between $\sim 1 - 100$~GeV$^2$~\cite{Armesto_2008}. The two regions cannot be described by a single formalism. Different approximations and extrapolations are needed in each case. 

\begin{figure}[t]
  \centering
   \includegraphics[width=1.0\linewidth]{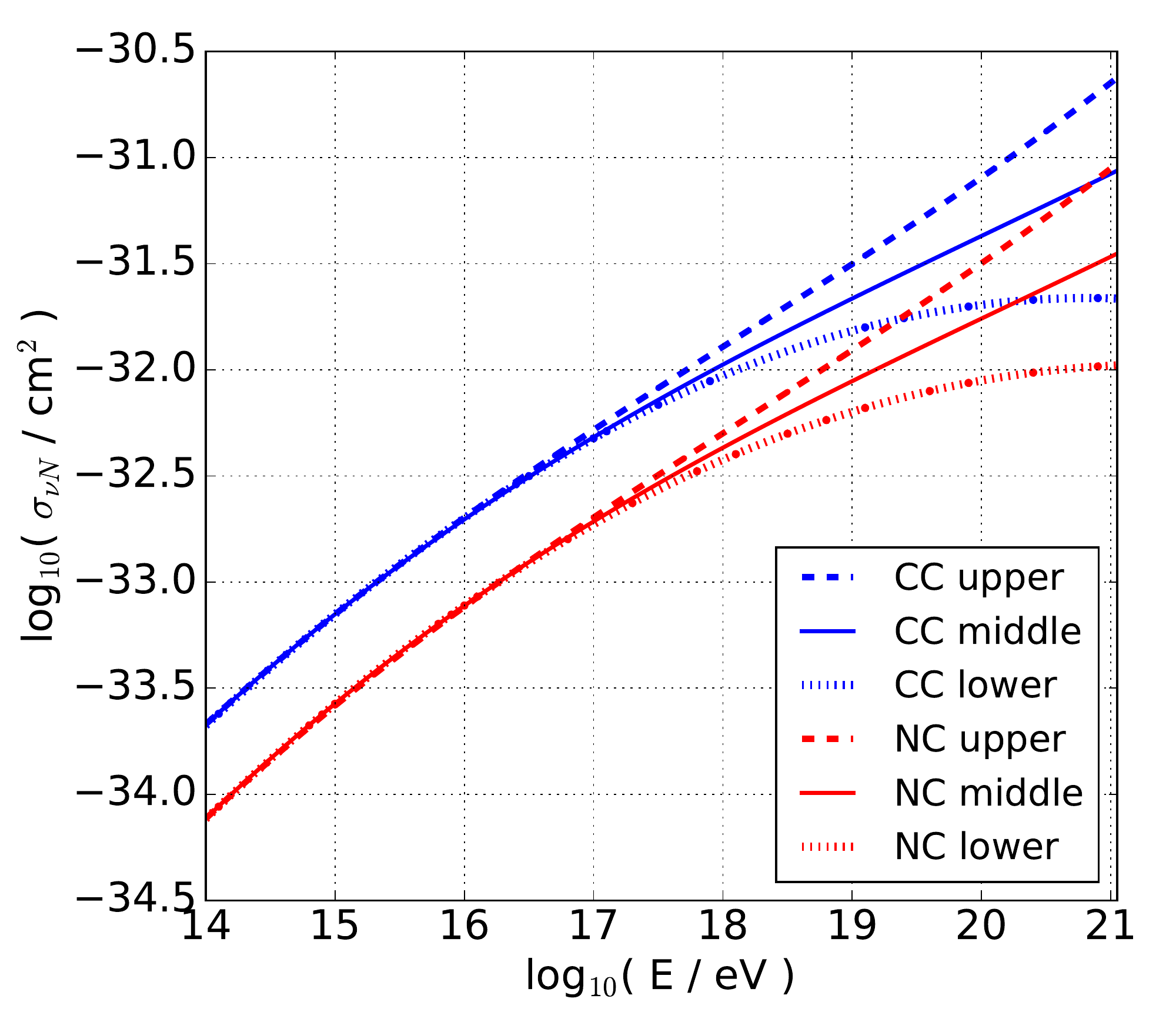} %
   \caption{Standard model charged current (CC) and neutral current (NC) neutrino-nucleon cross-sections $\sigma_{\nu N}$ as a function of energy from~\cite{Connolly_2011}. The upper and lower curves represent the upper and lower limits of the uncertainties due to the parton distribution function. In this work we consider the upper, middle, and lower curves in the calculation of the probability and energy distribution of exiting $\tau$ leptons.}
   \label{fig:cross_sections}
\end{figure}

\begin{table}[bp]
   \centering
   \begin{tabular}{l|cccc} 
     Cross-section     & $p_0$ & $p_1$ & $p_2$ & $p_3$ \\
    \hline
    CC upper   & -53.1 & 2.73 & -0.129 &   0.00237 \\
    CC middle  & -53.5 & 2.66 & -0.114 &   0.00182\\
    CC lower   & -42.6 & 0.49 & 0.003 &   -0.00133 \\
    NC upper   & -53.7 & 2.73 & -0.127 &   0.00231\\
    NC middle  & -54.1 & 2.65 & -0.112&    0.00175 \\
    NC lower   & -44.2 & 0.71 & 0.002 &   -0.00102 \\
   \end{tabular}
   \caption{Parameters for the log-log polynomial fit $\log_{10}(\sigma/{\rm cm}^2) = \sum_{k=0}^3p_k (\log_{10}(E/{\rm eV}))^k$ to the Standard Model neutrino cross-sections 
   (see Figure~\ref{fig:cross_sections}).}
   \label{tab:cross_sections}
\end{table}

\begin{figure}[t]
  \centering
   \includegraphics[width=1.0\linewidth]{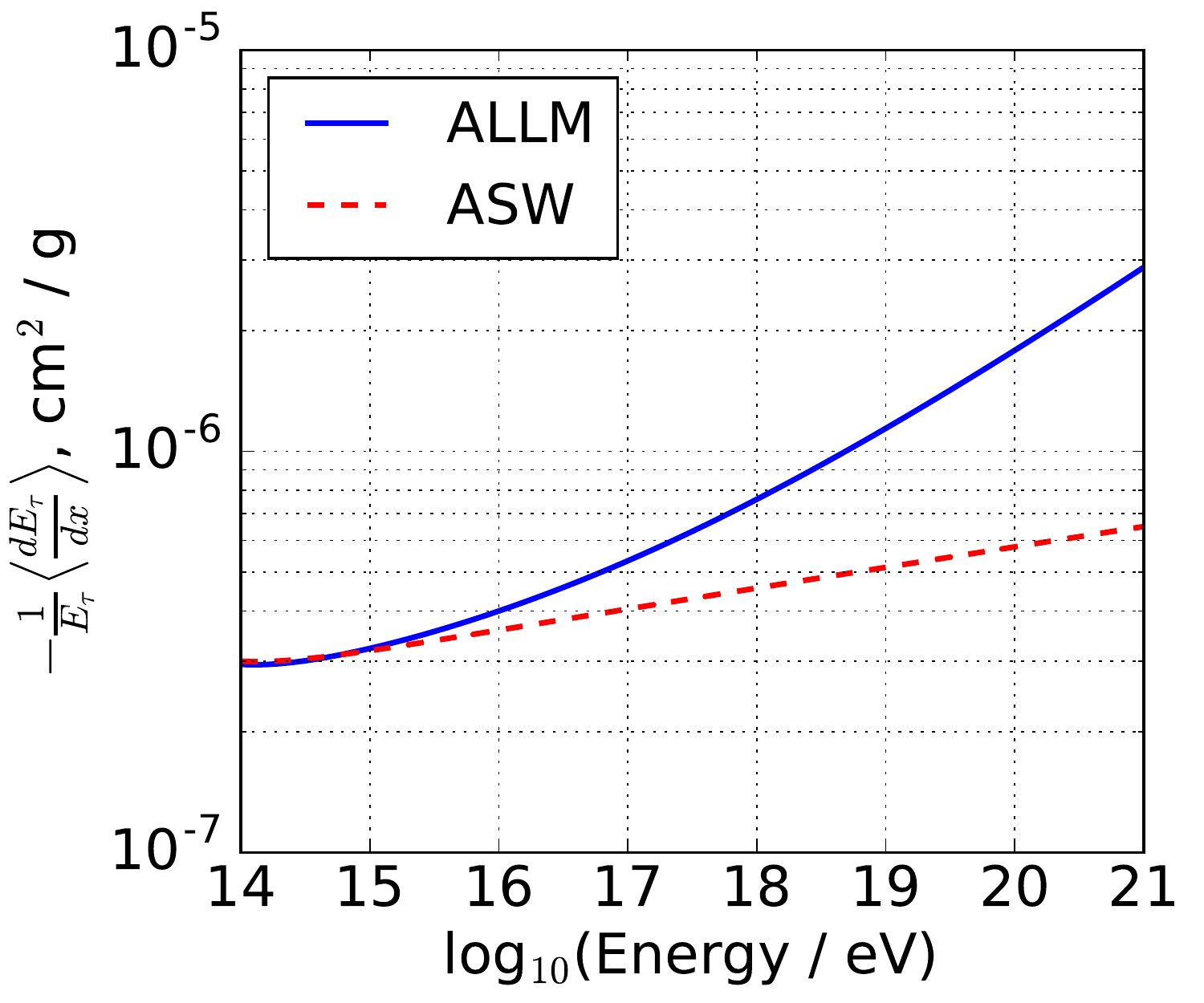} %
   \caption{Fractional $\tau$ lepton energy loss rate $\frac{1}{E_{\tau}} \left\langle \frac{dE_{\tau}}{dx}\right\rangle$, including ionization, pair production, bremsstrahlung, and photo-nuclear interactions as a function of energy. The two curves shown are for the ALLM~\cite{ALLM_97} and ASW~\cite{ASW_2005} photo-nuclear interaction models.}
   \label{fig:tau_E_loss}
\end{figure}

Standard model uncertainties in the neutrino-nucleon cross-section 
can be quite large reaching up to a factor of five at energies of $10^{21}$~eV,
according to ~\cite{Connolly_2011}. 
It is not the purpose of this work to explore these uncertainties. We have 
considered characteristic cross-sections as given in~\cite{Connolly_2011} 
(see Figure~\ref{fig:cross_sections} and Table~\ref{tab:cross_sections}).
We have chosen to include the central value of the cross
section together with the given parameterizations of the upper and lower 
limits of the uncertainty band as options. They give us an order of magnitude estimate 
of the uncertainties involved within the Standard Model. We note that
alternative calculations based on the dipole model~\cite{Jeong_2017} can give even slightly
lower values (order $20\%$) at $10^{18}$~eV.
Moreover there have been several studies on new physics that could suppress or 
enhance the cross-sections at high energies well beyond the chosen uncertainty 
band~(see for instance~\cite{Cornet_2001, Jain_2002, Reynoso_2013}). 

For this study, we have not considered the uncertainties in the $y$-distributions and we use the results from standard calculations using parton distributions~\cite{Lai_2000} throughout. The work of~\cite{Jeong_2017} included the full dependence on the $y$-distribution and concluded that the results can be reproduced accurately using only the average $\langle y \rangle$ as a function of energy. 
%
Between the models considered in~\cite{Jeong_2017}, changes in $\langle y \rangle$ range from -3\% to 7\% at $10^{15}$~eV and -15\% to 23\% at $10^{20}$~eV. 
The resulting fluxes in that reference show consistent results with different models within 25\%. If the logarithmic slope of the structure functions are significantly altered from the values obtained from extrapolating the parton distributions, the average value of $y$ can change~\cite{CastroPena:2000fx}. Dependence on the $y$-distribution are not addressed further in this article. Alternative parameterized cross-sections or $y$-distributions could be readily included in the simulation code to produce results for the $\tau$ lepton exit probability and energy distribution as a function of exit angle.

The $\tau$ lepton energy loss rate is often parameterized as,
\begin{equation}
\left\langle\frac{dE_{\tau}}{dx}\right\rangle = -a(E_\tau) - b(E_\tau)E_\tau.
\label{eqn:tau_energy_loss_rate}
\end{equation}
where $a$ and $b$ are slowly varying functions of the $\tau$-lepton energy. The parameter $a$ accounts for ionization processes and is practically constant 
$a\approx2\times10^{6}$~eV~cm$^2$/g 
at energies much larger than the $\tau$-lepton mass relevant to this work. The 
parameter $b(E_\tau)$ includes contributions from bremsstrahlung, pair production, 
and photo-nuclear interactions. 
At energies $E_{\tau}>10^{16}$~eV, photo-nuclear interactions have the largest
and most uncertain contribution to $b$, which is also due to uncertainties 
in the knowledge of the parton distribution functions in the 
corresponding kinematic region (see for instance \cite{Jeong_2017, Armesto_2008}).

We again consider parameterizations of two models of the 
contribution of the photo-nuclear 
interaction to the energy loss rate: 
ALLM~\cite{ALLM_97} and ASW~\cite{ASW_2005}. The fractional 
energy loss rate for each model is shown in Figure~\ref{fig:tau_E_loss}. For the purposes of 
this study, we take ALLM to be a typical model while ASW represents a model with suppressed 
photo-nuclear energy loss rate. The parameterizations are given by a power law in energy 
$b(E_\tau) = p_0+p_1(E_\tau/ \mbox{GeV})^{p_2}$ with parameters provided in
Table~\ref{tab:tau_E_loss}. In contrast to~\cite{Palomares-Ruiz_2006}, 
we use the same results for water and rock when the depth $x$ is expressed in
g~cm$^{-2}$ since differences due to nuclear effects are relatively
insignificant compared to other uncertainties. 


\begin{table}[htbp]
   \centering
   \begin{tabular}{l|rrc} 
     $b(E_{\tau})$    & $p_0$ (cm$^2$/g) & $p_1$ (cm$^2$/g)  & $p_2$ \\
    \hline
    ALLM  & 2.06$\times$10$^{-7}$ & 4.93$\times$10$^{-9}$ & 0.228 \\
    ASW  & -4.77$\times$10$^{-7}$ & 1.90$\times$10$^{-7}$ & 0.047 \\
   \end{tabular}
   \caption{Parameters for the power law fit to the $b(E_{\tau})$ function in to the energy loss rate in Equation~\ref{eqn:tau_energy_loss_rate} (see Figure~\ref{fig:tau_E_loss}). The parameterization is $b(E_\tau) = p_0+p_1(E_\tau/ \mbox{GeV})^{p_2}$ .}
   \label{tab:tau_E_loss}
\end{table}

The neutrino-nucleon cross-section and the photo-nuclear contribution to 
the $\tau$ lepton energy loss rate are calculated from extrapolated structure 
functions. 
We take the uncertainties in both processes to be independent (see however~\cite{Jeong_2017}). 
This approach is justified since the 
ranges of the kinematical variables ($Q^2$ and Bjorken-$x$)
that contribute to the neutrino cross-section 
(dominated by perturbative QCD) and $\tau$-lepton energy loss 
(where perturbative and non-perturbative effects are mixed), 
are quite different at $10^{18}$~eV energies, and none of the 
available parameterizations of the parton distributions are adequate 
to describe both ranges simultaneously~\cite{Armesto_2008}.

\section{Results}
\label{sec:results}
The simulation presented here allows for variations in emergence angle  $\bar\theta$, incident neutrino
energy $E_\nu$, thickness $D$ and density of the water layer, together with cross-section
and energy loss models as described above. The number of combinations are far 
too many for a comprehensive study. Instead, we will narrow down some 
comparisons that characterize the dependence on these variables. 
\subsection{Tau Lepton Production Dependence on Neutrino Energy}
We characterize the behavior of various injected neutrino energies while
taking the cross-sections corresponding to the middle curves
and the ALLM energy loss rate model. We treat the case of a $D=4$~km thick
layer of ice, corresponding to the thickest regions of the Antarctic ice
sheet. This is quite similar to what can be expected for an
average ocean depth of ($\sim3.7$~km) given that the density of ocean water is $\sim10\%$ greater than ice.

\begin{figure}[t]
  \centering
   \includegraphics[width=1.0\linewidth]{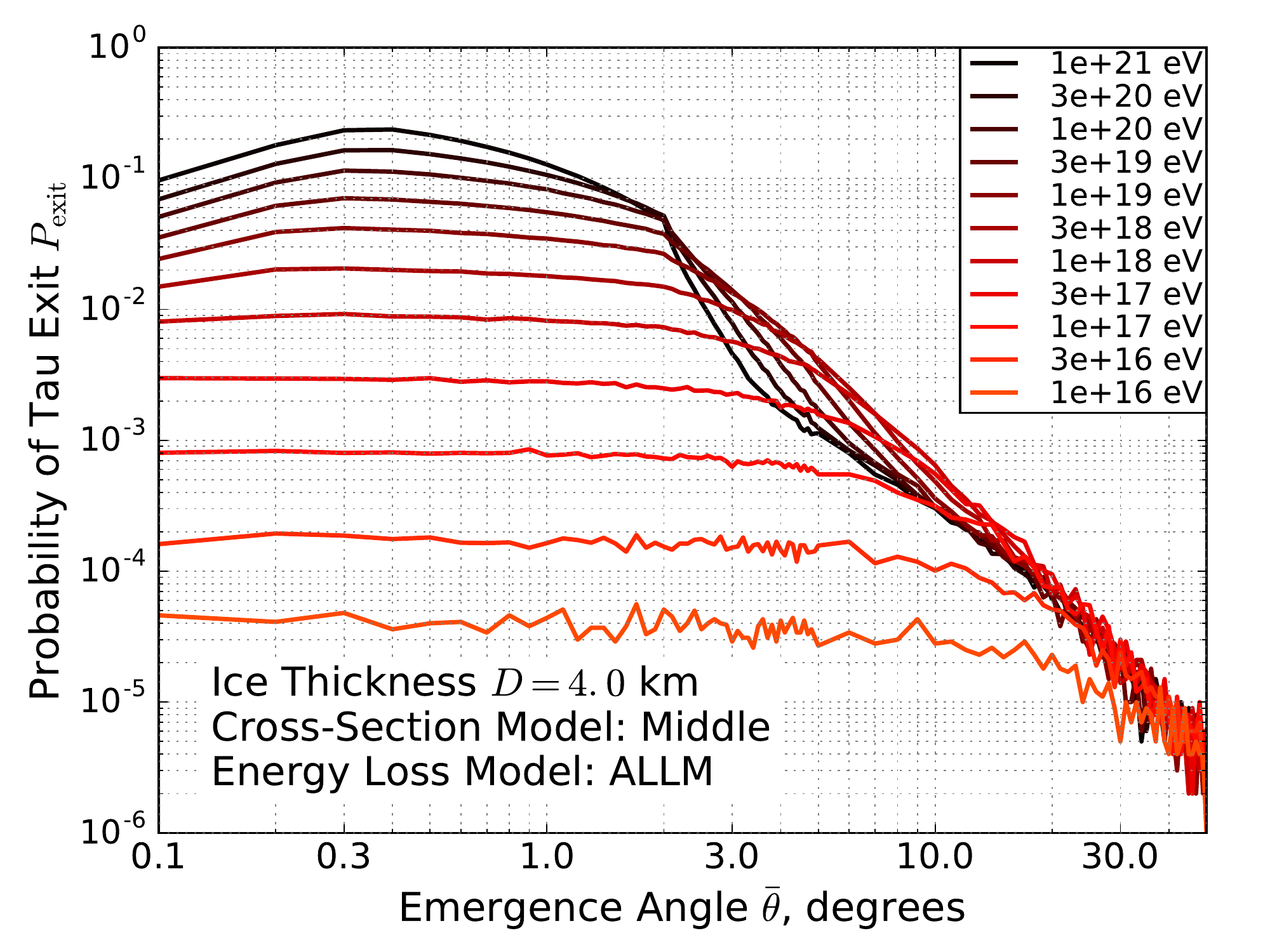} %
   \caption{The probability $P_\mathrm{exit}$ that a $\tau$ lepton exits the Earth's surface for emergence angles between 0.1$^{\circ}$ (Earth skimming) and 50$^{\circ}$ given a 4 km~thick layer of ice with standard cross-sections and energy-loss models. The feature at emergence angle of $2^{\circ}$ corresponds to the trajectory tangential to the rock layer beneath the 4~km thick layer of ice. }
   \label{fig:tau_exit_prob}
\end{figure}

\begin{figure}[!t]
  \centering
   \includegraphics[width=1.0\linewidth]{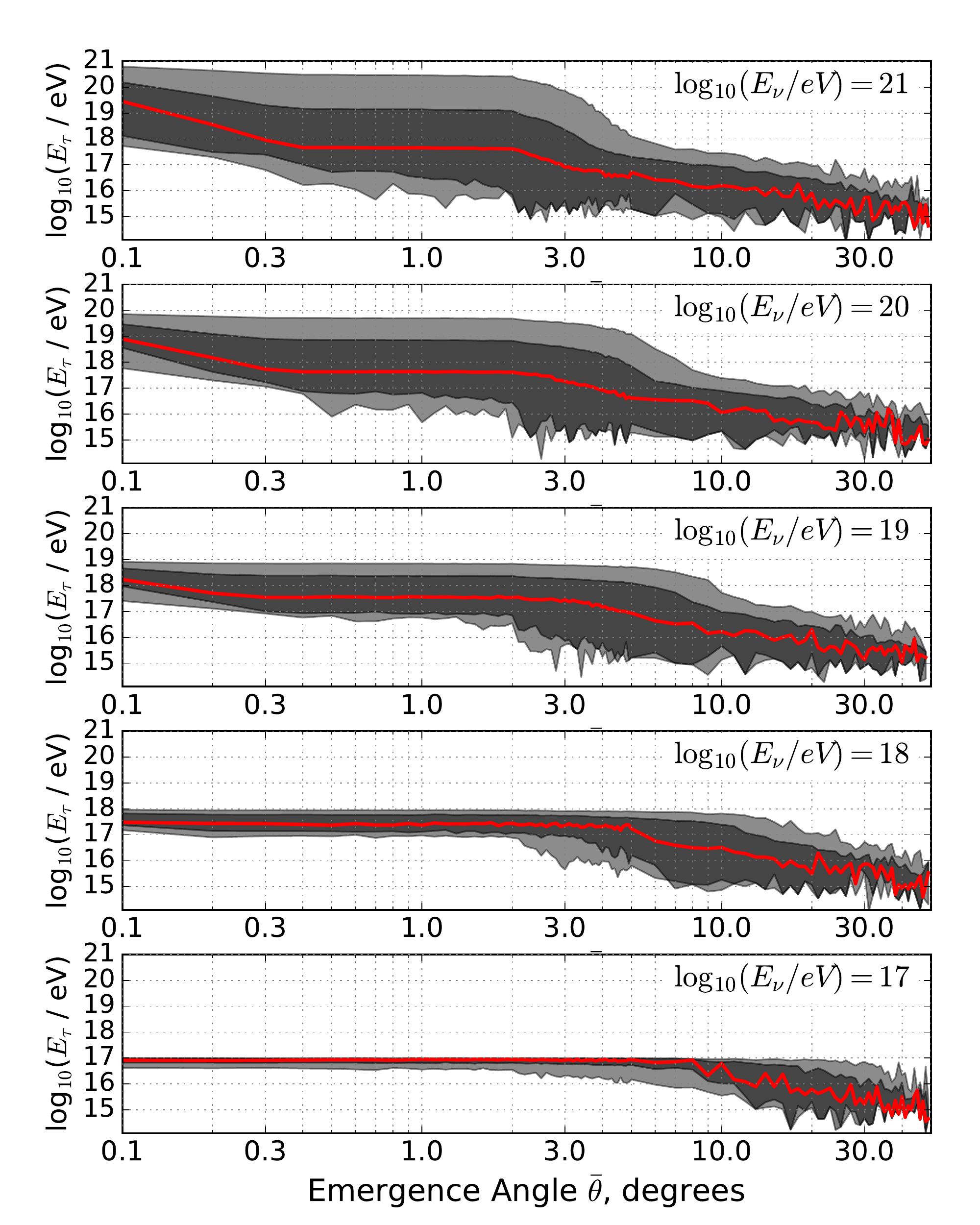} %
   \caption{The exiting $\tau$ lepton energies corresponding to some of the energies shown in Figure~\ref{fig:tau_exit_prob}. The red line shows the most probable exiting tau lepton energy. The dark (light) gray band shows the 68\% (95\%) densest probability interval. The features in the curves are caused by regions where various interaction processes processes dominate. See Figure~\ref{fig:tau_interactions} and text for details. }
   \label{fig:tau_energy_distrib}
\end{figure}

\begin{figure}[!t]
  \centering
   \includegraphics[width=1.0\linewidth]{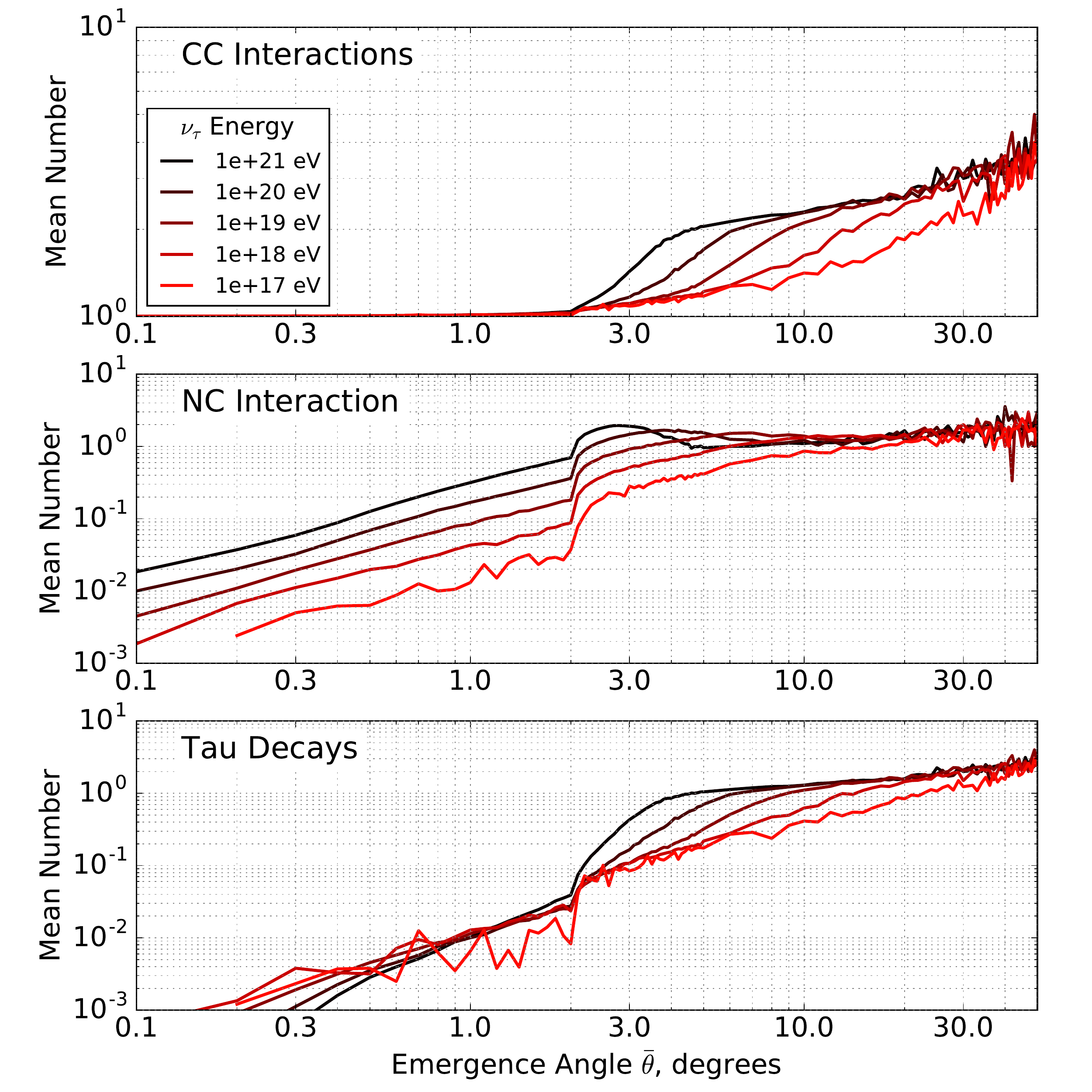} %
   \caption{The mean number of CC, NC interactions, and tau lepton decays as a
     function of emergence angle for various incident neutrino energies
     corresponding to
     Figures~\ref{fig:tau_exit_prob}~\&~\ref{fig:tau_energy_distrib}. Top
     panel: the mean number of CC interaction must be at least one since we
     are selecting for particles resulting in a $\tau$ lepton exiting the
     Earth's surface. Middle Panel: The mean number of neutral current
     interactions. The sharp transition at emergence angle $\bar \theta=2^{\circ}$ corresponds
     to the direction tangential to the subsurface rock beneath 4~km thick
     layer of ice. Bottom panel: the mean number of $\tau$ lepton decays also
     show a feature at $\bar\theta=2^{\circ}$. Note that for $\bar\theta<2^{\circ}$ the particle traverse ice only while for $\bar\theta>2^{\circ}$ the particle traverses a combination of rock and ice, which affects the behavior of $\tau$ lepton and $\nu_\tau$ interactions.}
   \label{fig:tau_interactions}
\end{figure}

If a $\nu_\tau$ enters the Earth either a  $\tau$ lepton or a
$\nu_\tau$ must exit at the other side because the $\tau$-lepton always
decays into a $\nu_\tau$ even if the energy of the exiting lepton can be
significantly reduced. 
For the $\tau$ lepton to be injected to the atmosphere
the CC interaction must happen before the exit point and within the tau 
range. The closer it happens to the exit point the higher the tau 
energy at injection. The effective interaction volume for neutrino 
interactions is adjacent to the Earth surface and scales with the $\tau$ lepton range.

In Figure~\ref{fig:tau_exit_prob} we show the
probability that a $\tau$ lepton exits the volume of the Earth
($P_\mathrm{exit}$) given an injected $\nu_\tau$ of energy $E_{\nu}$. The
probability is highest for large $E_{\nu}$ in Earth-skimming trajectories at
small emergence angles ($\bar\theta<2^{\circ} 
$). The feature at $\bar{\theta}=2^{\circ}$ 
corresponds to the direction of propagation that is tangential to the layer of rock beneath the 
D=4~km thick ice layer (see Figure~\ref{fig:tau_geom}). For
lower $E_{\nu}$, when the $\nu_\tau$ interaction cross-section 
is reduced, $P_\mathrm{exit}$ is typically smaller and the curves in 
Figure~\ref{fig:tau_exit_prob} broaden and have less features.

The emerging $\tau$ lepton energy $E_{\tau}$ has a distribution 
which depends on the emergence angle and on the primary neutrino energy in a
complex way as 
shown in Figure~\ref{fig:tau_energy_distrib}. In each panel we show the mode of the $E_{\tau}$ distribution along with the densest ranges containing 68\% and 95\% 
of the distribution as dark and light-gray areas, respectively.
In Figure~\ref{fig:tau_interactions} we also 
show the mean number of $\nu_\tau$ CC and NC interactions as well as 
the mean number of $\tau$ lepton decays as a function of $\bar{\theta}$ 
for selected neutrino energies,  
to better characterize the features in Figure~\ref{fig:tau_energy_distrib}.

The behavior of the conversion probabilities to exiting $\tau$ leptons and the 
corresponding $E_\tau$ distributions can be understood in terms of  three scales. 
The length of the Earth's chord is linearly dependent on $\sin \bar{\theta}$,  being 24~km (240~km) for $0.1^\circ$ ($1^\circ$).
Its grammage scales with length up to $\bar\theta \simeq 2^\circ$ where there is a discontinuity tangent to the bedrock. 
Its length is relevant in relation to the $\tau$-lepton decay while its grammage governs neutrino interactions and $\tau$-lepton energy 
loss. A second scale is the interaction depth for the neutrino 
conversion to a $\tau$ lepton, which decreases as the neutrino energy rises. 
For the reference cross-section it corresponds to an interaction length of
about 1800~km in ice for $10^{18}$~eV neutrinos and decreases by about a 
factor of two per decade becoming about 225 km of ice for $10^{21}$~eV. 
As long as the interaction depth is much greater than that of the chord, we can 
expect a single CC interaction and negligible effects to the neutrino
flux because of attenuation or regeneration. 
The third scale is given by the $\tau$-lepton range. At low energies it
scales with the $\tau$-lepton energy but when the energy exceeds a critical value,  
$E_\mathrm{crit}$, the $\tau$-lepton range is determined by energy loss and can be 
shown to scale roughly as log$(E_\tau/E_\mathrm{crit})/b(E_\tau)$. In this regime the $\tau$-lepton loses 
energy rapidly until it approaches $\sim E_\mathrm{crit}$ when decay 
quickly follows. If $b(E_\tau)$ and the target density $\rho$ are constant, 
$E_\mathrm{crit} \sim m_\tau c^2/(\rho c \tau_\tau~b)$, where $m_\tau$ and
$\tau_\tau$ are the rest mass and lifetime of the $\tau$-lepton~\cite{Zas_2005}. 
The dependence on $\rho$ is responsible for some of the differences between ice and rock targets. For $\rho=0.925$~gcm$^{-2}$ and $b\sim 6.5~10^{-7}$~g \ cm$^{-2}$ its value is $E_\mathrm{crit} \sim
3.4\times10^{17}$~eV. 

We can see a distinct behavior of the $E_\tau$ distributions in different 
$\bar\theta$ ranges. 
The transition points depend on the neutrino energy. 
We consider two broad cases in relation to neutrino energy:

(1) 
For $E_{\nu} \gtrsim 3~10^{18}$~eV
the trajectory undergoes, on average, a single CC interaction 
up to about $2^\circ$, as seen in
Figure~\ref{fig:tau_interactions}, 
and the $\tau$ lepton can lose a large fraction of its energy before exiting. 
For increasing $\bar\theta$, the trajectory has higher interaction 
probability and more $\tau$-lepton energy loss on average (see Figure~\ref{fig:tau_energy_distrib}) and as a result the mode 
of $E_\tau$ drops and its variance increases. 
The conversion probability peaks when the Earth's chord roughly matches the tau 
range at $\sim 0.4^\circ$ ($\sim 0.2^\circ$) for $E_\nu=10^{21}$
($10^{19}$)~eV. 
At larger $\bar \theta$ the neutrino can interact early in the chord 
so that the $\tau$-lepton decays before exiting, decreasing the conversion probability. 

The mode of the $E_\tau$ distribution has a roughly exponential 
decrease as $\bar\theta$ increases until it reaches a plateau when the
chord matches the $\tau$-lepton range (see Figure~\ref{fig:tau_energy_distrib}). 
In the plateau the increase of grammage as $\bar\theta$ rises hardly alters 
the exiting $\tau$-lepton energy distribution, the relevant condition being that the interaction is produced
within the $\tau$-lepton range of the exiting point. The average emerging $\tau$-lepton energy  
is roughly $\sim E_\mathrm{crit}$ in the plateau for all neutrino energies since
this is the energy at which the tau preferentially decays. The plateau ends 
when the Earth's chord reaches a grammage that attenuates the incident 
neutrino flux.
For larger $\bar\theta$ regeneration becomes quite important as
can be seen in Figure~\ref{fig:tau_interactions}.  

(2) For lower neutrino energies, $E_{\nu} \leq 3\times10^{18}$~eV, the
characteristics are similar but the transitions occur at very different values 
of $\bar\theta$. The increase in the conversion probability 
as $\bar\theta$ rises happens at extremely low angles that can hardly
be noticed in the plots shown because the $\tau$-lepton range is very much reduced. 
The plateau reaches larger values of $\bar\theta$ because 
attenuation and regeneration of the neutrino flux require more matter depth.

The relatively large variance of $E_\tau$ at high neutrino energies and low $\bar\theta$ is largely due to the fact that the neutrino CC interaction can occur anywhere in the trajectory. 
The upper end of the $E_\tau$ distribution is constrained by the neutrino energy and the lower tail by
energy loss and decay of the $\tau$~lepton. 
As the neutrino energy drops the variance of the $E_\tau$ distribution reduces. 
At low incident neutrino energies the tau energy loss is relatively small because decay dominates. 
On the other hand when $\bar\theta$ is sufficiently large for regeneration to become important, the width of the $E_\tau$ distribution becomes again relatively wide because of the fluctuations inherent to the regeneration processes.

\subsection{Dependence on Regeneration}


\begin{figure}[!t]
  \centering
   \includegraphics[width=1.0\linewidth]{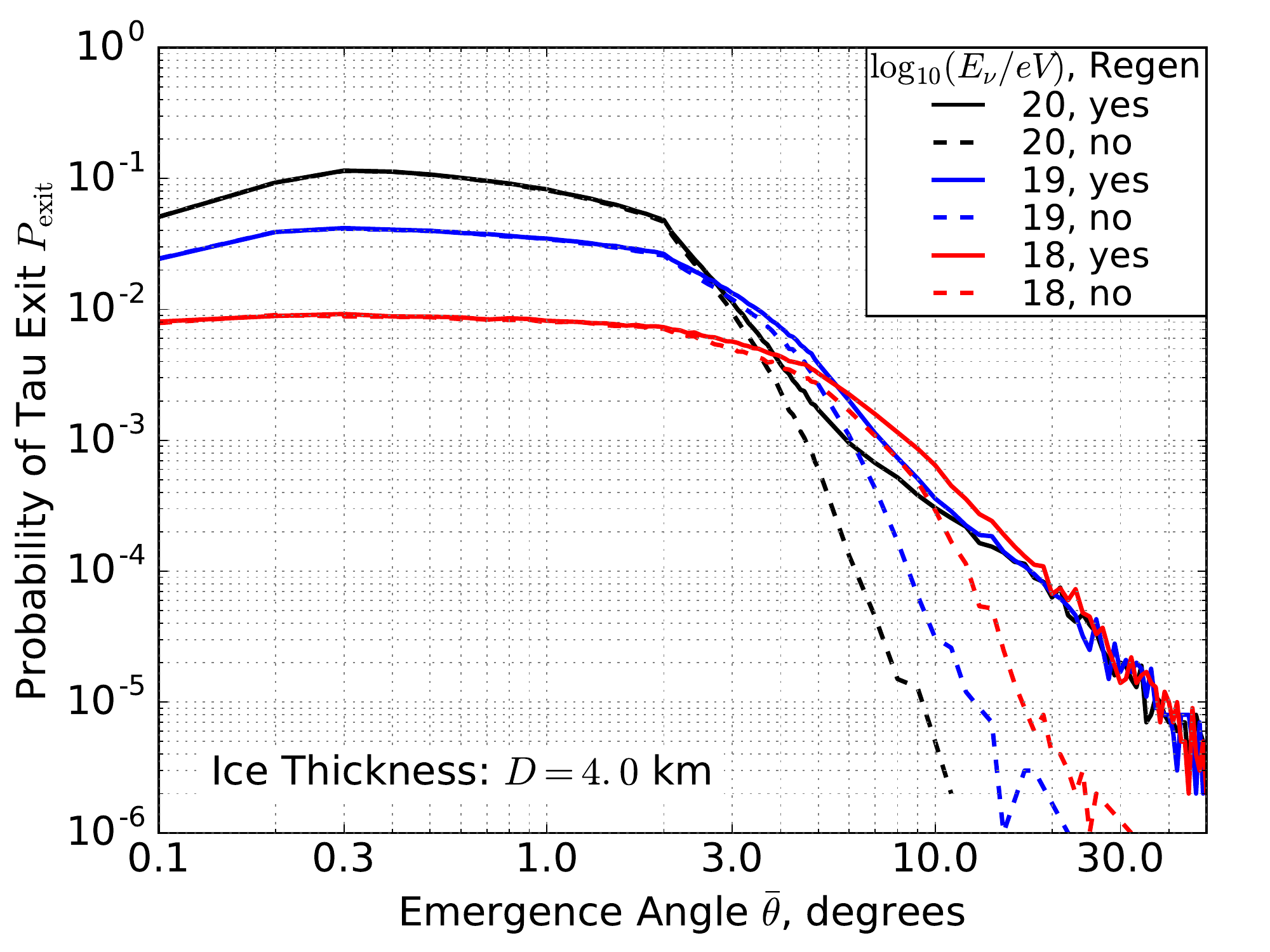} %
   \caption{The probability that a $\tau$ lepton exits Earth's surface including and excluding the effect of $\nu_\tau$ regeneration given a 4~km thick layer of ice and standard neutrino cross-section and tau lepton energy-loss models. Excluding regeneration significantly underestimates the probability of exiting $\tau$ leptons for $\bar\theta>2^{\circ}$, where the trajectories propagate through rock rather than pure ice. }
   \label{fig:tau_exit_regen}
\end{figure}

\begin{figure}[!t]
  \centering
   \includegraphics[width=1.0\linewidth]{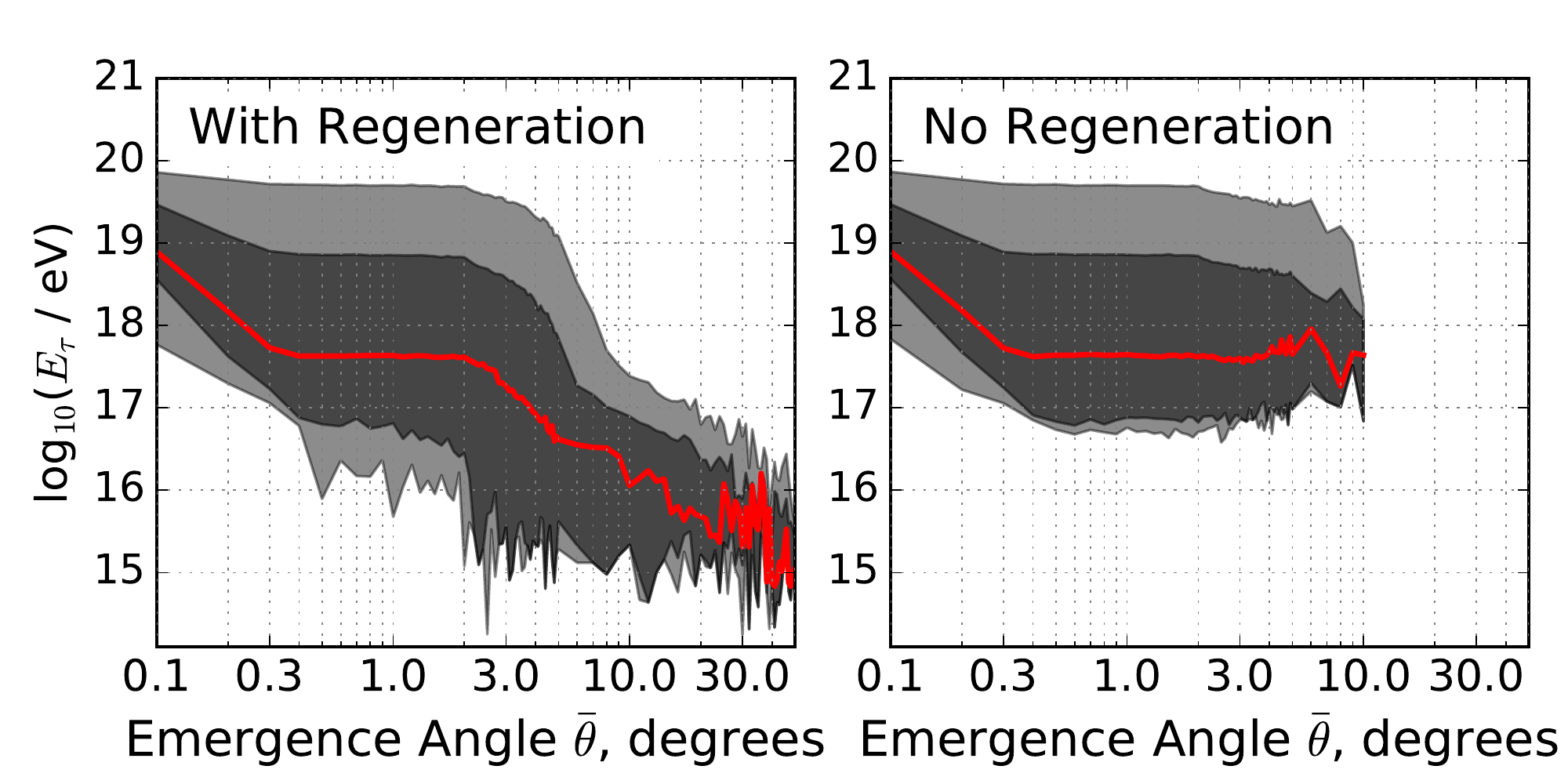} %
   \caption{The exiting $\tau$ lepton energies corresponding to $E_{\nu}=10^{20}$~eV in Figure~\ref{fig:tau_exit_regen} with and without regeneration. Excluding regeneration suppresses exiting $\tau$ leptons with energy $E_{\tau}<10^{17}$~eV.} 
   \label{fig:tau_energy_distrib_regen}
\end{figure}

To illustrate the magnitude and importance of the effect of including the
$\nu_\tau$ regeneration we compare simulation results with regeneration
deactivated. In Figure~\ref{fig:tau_exit_regen} we show the $\tau$ lepton exit
probability as a function of emergence angle with and without regeneration. In
the absence of regeneration, the exit probability is clearly suppressed for
emergence angles $\bar{\theta}> 2^{\circ}$, corresponding to the exit angles
where the particle traverses rock as opposed to pure ice. With increasing
emergence angles, ignoring $\nu_\tau$ regeneration underestimates the $\tau$ lepton exit probability by orders of magnitude.

The $\tau$-lepton exit energy distributions for injected neutrino energy of
$E_{\nu}=10^{20}$~eV with and without regeneration are shown
in Figure~\ref{fig:tau_energy_distrib_regen}. Neglecting
$\nu_\tau$ regeneration is consistent with regeneration while
$\bar\theta<0.4^{\circ}$. In the range $0.4^{\circ}<\bar\theta<3^{\circ}$ the
most probable values are similar, however, the low energy tail
of the $E_{\tau}$ distribution is missing in this approximation. For
$\bar\theta>3^{\circ}$ the energy distributions vary significantly. This is to
be expected given that the number of $\tau$-lepton decays becomes significant in this
range as shown in Figure~\ref{fig:tau_interactions}. At
$\bar\theta>10^{\circ}$ the distribution is not shown 
in Figure~\ref{fig:tau_energy_distrib_regen} due to a lack of events.

\subsection{Dependence on Ice Layer Thickness}
\begin{figure}[t!]
  \centering
   \includegraphics[trim = {0.2cm 1.0cm 1.0cm 0.2cm}, clip, width=0.95\linewidth]{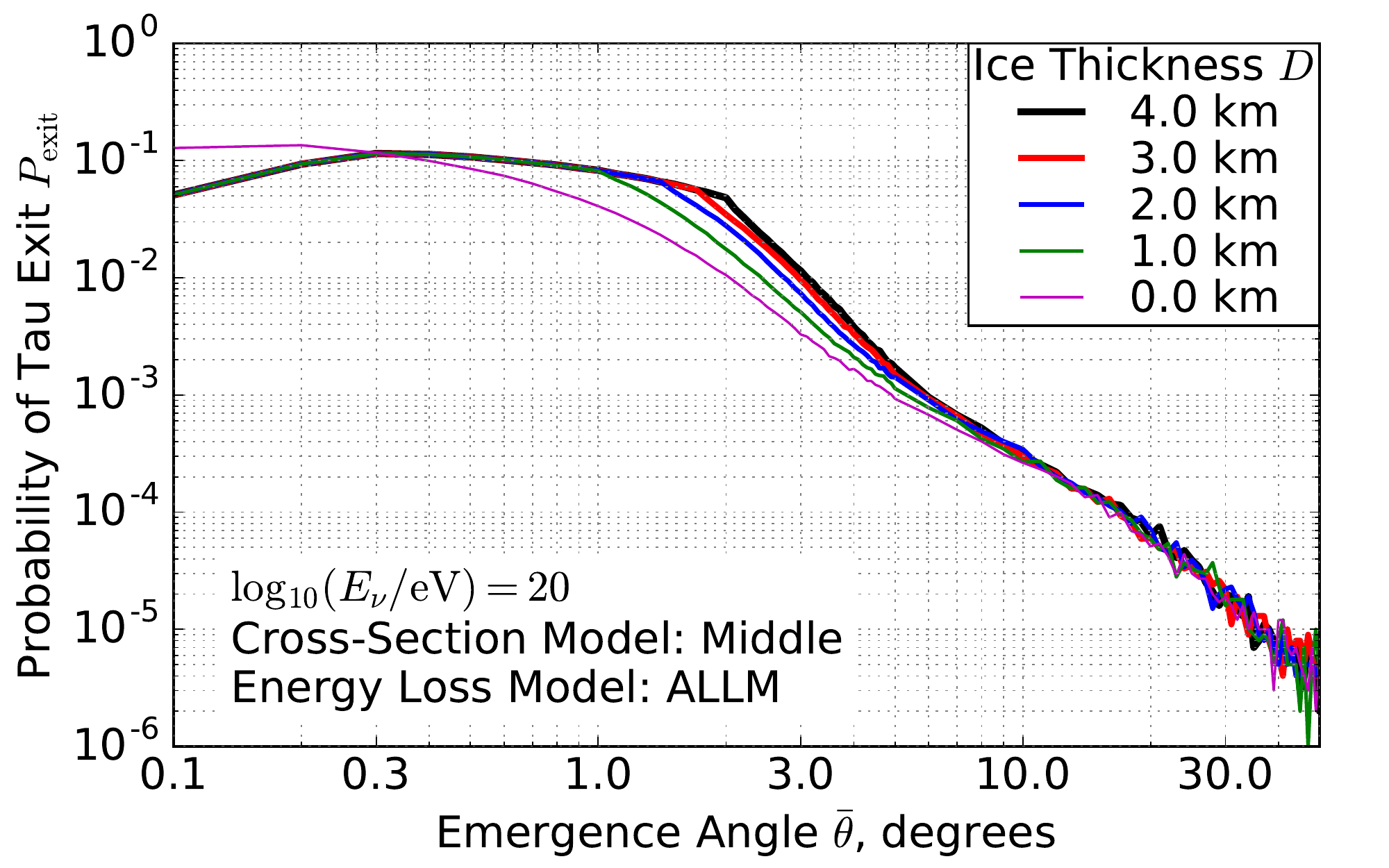} 
   \includegraphics[trim = {0.2cm 1.0cm 1.0cm 0.2cm}, clip, width=0.95\linewidth]{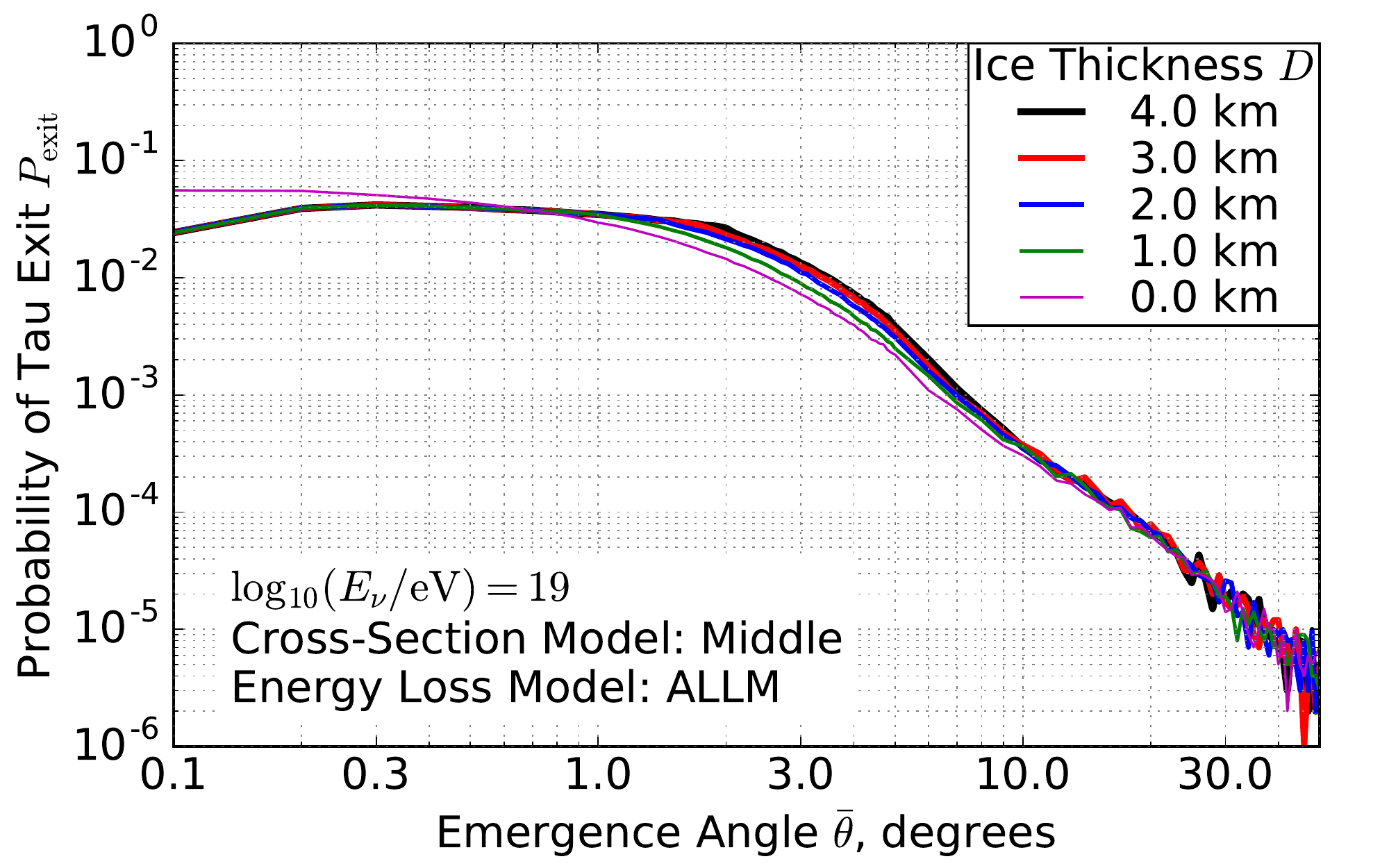} 
   \includegraphics[trim = {0.2cm 1.0cm 1.0cm 0.2cm}, clip, width=0.95\linewidth]{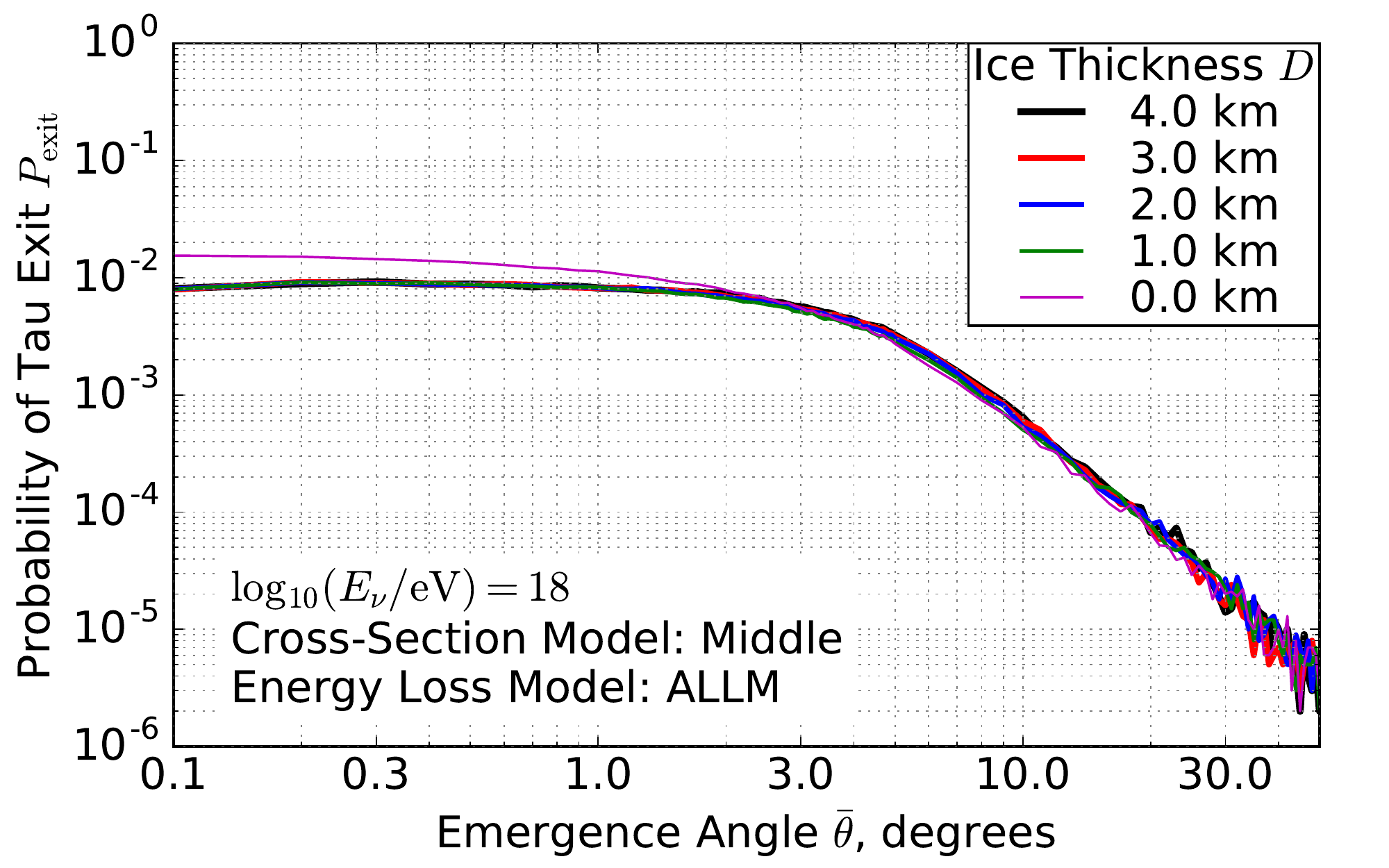} 
   \includegraphics[trim = {0.2cm 0.1cm 1.0cm 0.2cm}, clip, width=0.95\linewidth]{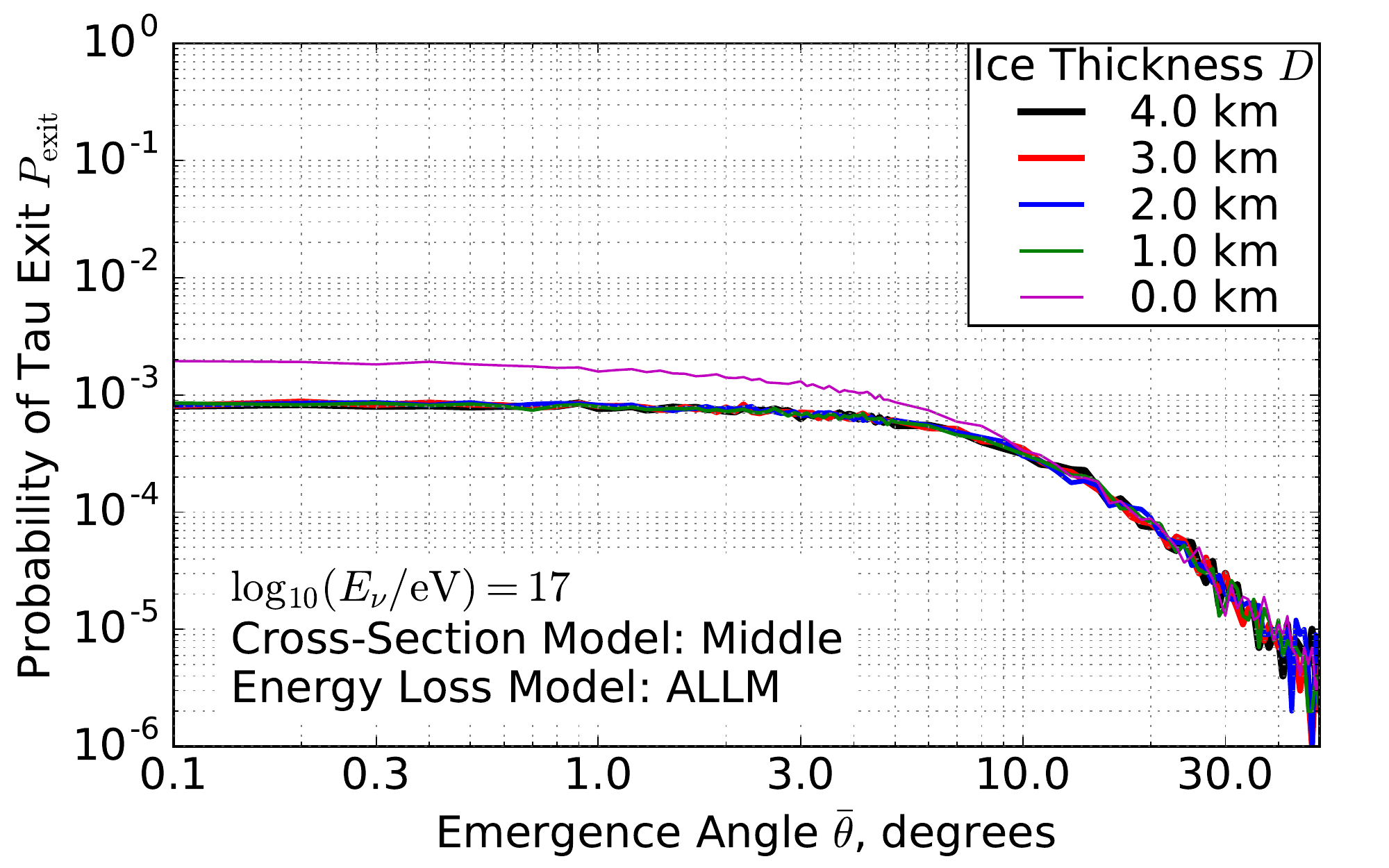} 
   \caption{The probability that a $\tau$ lepton exits the Earth's surface for various energies and ice thicknesses, including bare rock, assuming standard cross-section and energy-loss models. From top to bottom, the input neutrino energies are 10$^{20}$, 10$^{19}$, 10$^{18}$, and 10$^{17}$~eV. A layer of ice is favorable to exiting $\tau$ leptons for neutrino energies $>10^{18}$~eV while bare rock is favorable for neutrino energies $<10^{18}$~eV. See text for details.
\vspace{-30pt}
}
   \label{fig:tau_exit_prob_depth}
\end{figure}

The presence of a layer of ice or water can be an important consideration for the design of experiments. In Figure~\ref{fig:tau_exit_prob_depth} we show the dependence of $P_\mathrm{exit}$ on the ice thickness for several neutrino energies between 10$^{17}$--10$^{20}$~eV. For energies $>10^{18}$~eV, there is a stark difference between bare rock and the presence of a layer of ice, even if just 1~km thick. For each thickness, a break in the curve is found at the angle corresponding to the direction of propagation that is tangential to the layer of rock beneath the ice 
(1$^\circ$, 1.4$^\circ$, 1.8$^\circ$, 2.0$^\circ$ for 1, 2, 3, and 4 km ice thickness, respectively). For $\bar\theta$ around the angle at which this feature occurs, the exit probability can be close to a factor of five greater than bare rock at $E_\nu=10^{20}$~eV, in qualitative agreement with the results of~\cite{Palomares-Ruiz_2006}. At neutrino energies $E_\nu<10^{18}$~eV, bare rock has an advantage over the presence of a layer of ice. For $E_\nu=10^{17}$~eV, bare rock has roughly twice the $P_\mathrm{exit}$ than cases with an ice layer for emergence angles $\bar{\theta}<3^{\circ}$.

The effect is not intuitive because the relative probabilities that a $\tau$ lepton exits the atmosphere depend on neutrino energy. 
The results for neutrino energies of $10^{19}$~eV and $10^{20}$~eV shown in Figure~\ref{fig:tau_exit_prob_depth} are characterized by the $\tau$ lepton being above the critical energy $E_\mathrm{crit}$ described in section IV-A while $10^{17}$~eV and $10^{18}$~eV are below it. 
E$_\mathrm{crit}$ effectively separates the energy into high and low energy regions in which the $\tau$-lepton range is respectively governed by energy loss or decay. 
For neutrino energies $10^{19}$~eV and $10^{20}$~eV, the $\tau$-lepton range is dominated by energy loss while for $10^{17}$~eV and $10^{18}$~eV, the range is dominated by decay lifetime.

In the $10^{20}$~eV regime the $\tau$-lepton range is of order 60 km in ice and 25 km in rock for the ALLM model.
If the chord length is larger than this range, the region where the $\tau$ can exit efficiently will be proportional to the $\tau$ range and independent of the chord length. 
This can be seen in Figure~\ref{fig:tau_exit_prob_depth} for rock at $E_\nu=10^{20}$~eV above $0.1^\circ$ (22 km of chord length) and below $\sim0.2^\circ$. 
In this small angular range and for this energy, the larger interaction probability of $\nu_\tau$ in rock makes the exiting probability larger than in ice. 
However, at this same energy and for $\bar\theta>0.3^\circ$, where the range in rock is smaller than the chord, the lower density of ice reduces the $\tau$-lepton energy loss, and as a consequence the $\tau$-lepton range and hence the exit probability are larger.
As the emergence angle $\bar\theta$ (and chord length) increases, the chord (150~km at $0.7^\circ$ for $10^{20}$ eV) becomes comparable to the neutrino interaction length and attenuation sets in suppressing the exit probability.
%
%

In the $10^{17}$~eV regime, on the other hand, the $\tau$-lepton range is of order 5 km. 
At this energy regime, there is no significant difference in decay range for rock and ice because energy loss (and therefore the density) is not a significant effect. 
However, the higher density of rock increases the probability that a $\nu_\tau$ will interact near the surface compared to ice, which explains the higher value of $P_\mathrm{exit}$ in the bottom graph of Figure~\ref{fig:tau_exit_prob_depth}.
At high energies ($10^{19}$~eV and above), the curves show a discontinuity when the trajectory is tangent to the rock surface above which neutrino attenuation in the rock takes over while at low energies there is no discontinuity. 
This effect is also dependent on whether the $\tau$-lepton range is dominated by energy loss or decay lifetime since the former is sensitive to the density of the medium while the latter is not.

Finally, there is a change of behavior when regeneration takes over at angles of order $\bar\theta > 3^\circ$ for $E_\nu=10^{20}$~eV and $\bar\theta > 10^\circ$ for $E_\nu=10^{17}$~eV. 
%
%
In this regime the relative change in grammage for the two cases is small and the curves for rock and ice become very similar.

As a general conclusion, the increased $P_\mathrm{exit}$ and exiting $E_{\tau}$ indicates that the presence of an ice or water layer is highly favorable for detection at energies $E_\nu>10^{18}$~eV while bare rock is more favorable for $E_\nu< 10^{18}$~eV.

\subsection{Model Dependence}
\begin{figure}[t]
  \centering
   \includegraphics[width=1.0\linewidth]{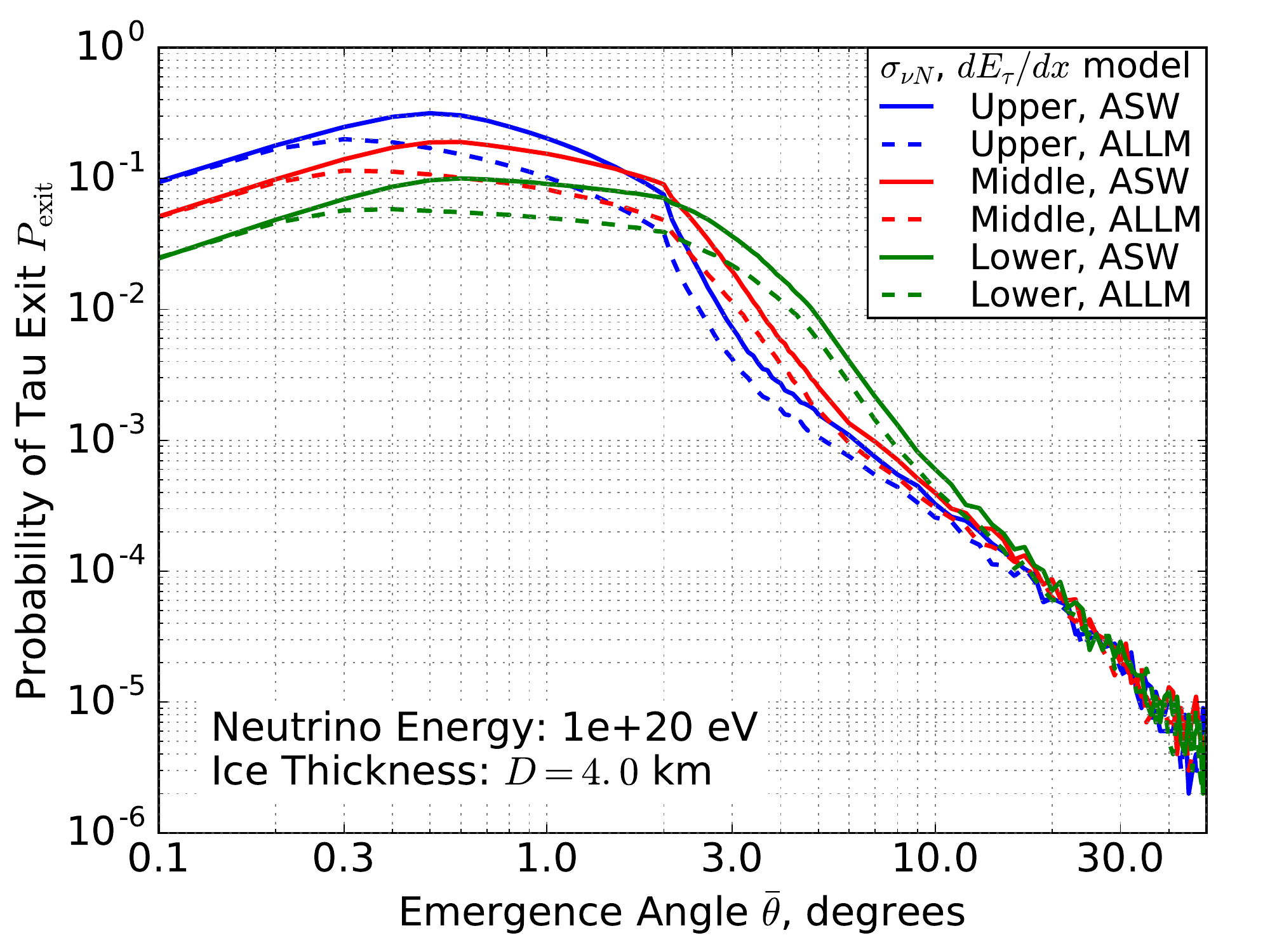} %
   \caption{The probability that a $\tau$ lepton exits the Earth's surface for various combinations of neutrino cross-section and $\tau$ lepton energy-loss models given a 4~km thick ice layer for a $10^{20}$~eV injected $\nu_\tau$. Lowering the cross-section has the general effect of reducing the $\tau$ lepton exit probability for emergence angles below where the trajectory is tangential to the subsurface rock layer while increasing the probability for larger emergence angles. The ASW energy loss rate model, which is suppressed compared to the more standard ALLM model, results in an overall increase $\tau$ lepton exit probability.}
   \label{fig:tau_exit_prob_models}
\end{figure}

We have considered different neutrino cross-section models consistent with the uncertainties in the standard model and a couple of $\tau$ lepton energy loss rate models to illustrate the effect of suppression. 
In Figure~\ref{fig:tau_exit_prob_models} we show the $P_\mathrm{exit}$ curves corresponding to combinations of these models. 
The effect of reducing the cross-section is that $P_\mathrm{exit}$ decreases below an emergence angle that coincides with the direction tangential to the bedrock ($\bar\theta_C\sim2^\circ$), for the chosen ice layer depth of 4 km, while it increases at larger emergence angles. 
This is because for low grammage the interaction probability drops while in the high grammage region the attenuation of the $\nu_\tau$ flux is reduced for a smaller cross-section. 
The values of $P_\mathrm{exit}$ are higher for the saturated ASW model with lower energy loss than for the more standard ALLM model.  
Differences can reach up to a factor of ten in the region of $\bar\theta \sim 4^\circ-5^\circ$.

\begin{figure}[t]
  \centering
   \includegraphics[width=1.0\linewidth]{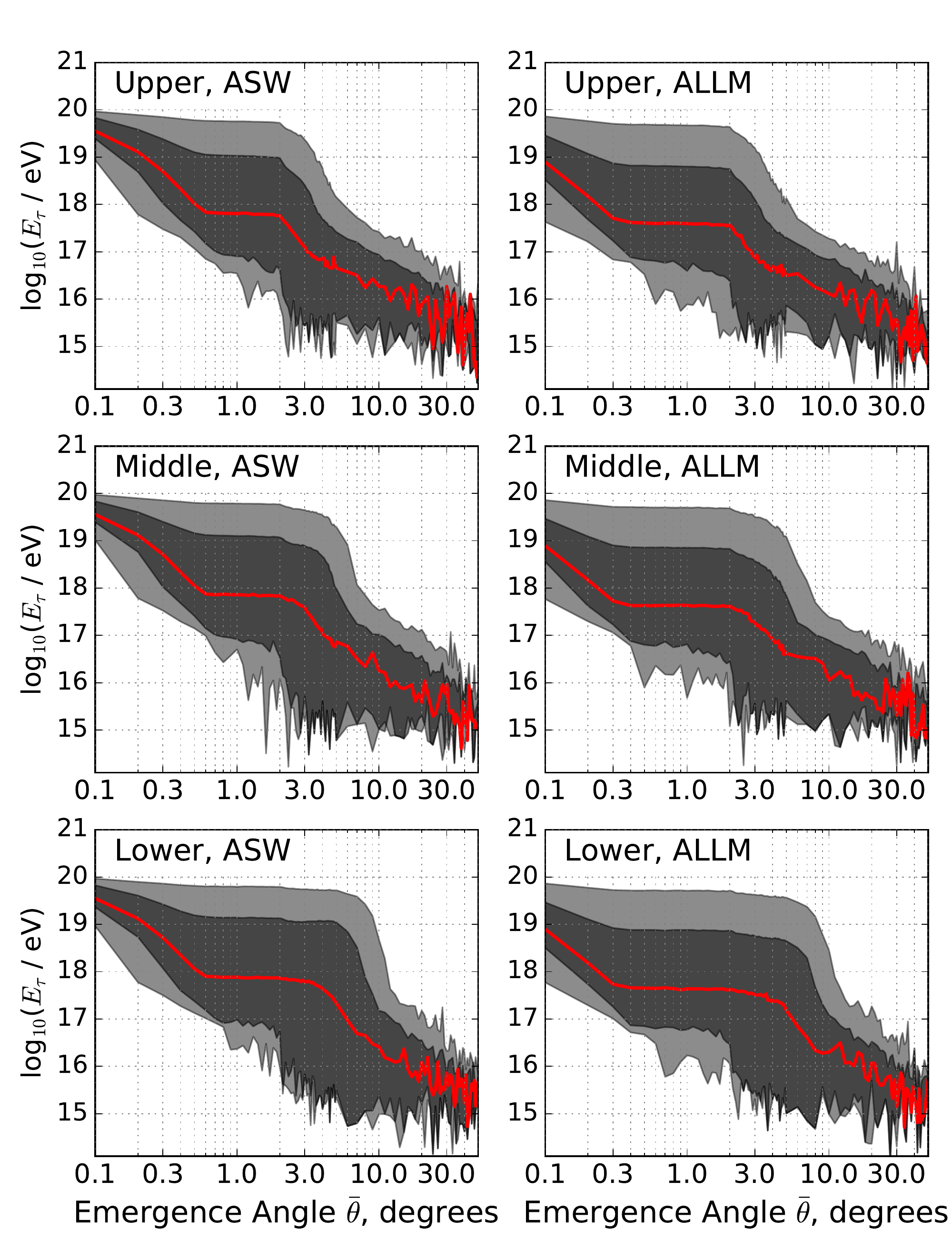} %
   \caption{The exiting $\tau$ lepton energies for various models corresponding to Figure~\ref{fig:tau_exit_prob_models}. On each panel, the cross-section model and energy loss rate models are labeled on the top right corner. The variance in exiting $\tau$ lepton energies tends to increase as the cross-section increases for trajectories that traverse mostly rock. The energy loss model changes the range of emergence angles where the most probable energies plateaus. }
   \label{fig:tau_energy_models}
\end{figure}

All curves converge for emergence angles $\bar\theta>20^{\circ}$ where the
effect of regeneration is strongest. Regeneration has the
  effect of reducing the neutrino energy until it reaches a value at which it
  is not likely to interact. It can be understood as an {\it energy shift} 
  to a fixed average energy that decreases as $\bar\theta$ increases but is
  independent of the primary neutrino energy. At $\bar\theta>20^{\circ}$ the
  energy is shifted down to energies below $10^{16}$~eV where there is 
hardly any difference between the considered models. 

The corresponding $\tau$-lepton $P_\mathrm{exit}$ energy distributions are shown in Figure~\ref{fig:tau_energy_models}. 
Decreasing $\tau$-lepton energy loss tends to narrow the distributions of $E_\tau$, increasing the mode of the distribution in the region of small $\bar\theta$ and increasing the angle at which the plateau in $E_\tau$ starts. 
Decreasing the neutrino cross-section tends to increase the angle at which the plateau ends and regeneration starts to dominate. 
This behavior indicates that with sufficient statistics, energy range, 
and range of emergence angles, the neutrino cross-section and $\tau$-lepton 
energy loss rate models could be constrained.

\subsection{Exiting Tau Fluxes}
\begin{figure}[!t]
  \centering
   \includegraphics[width=1.0\linewidth]{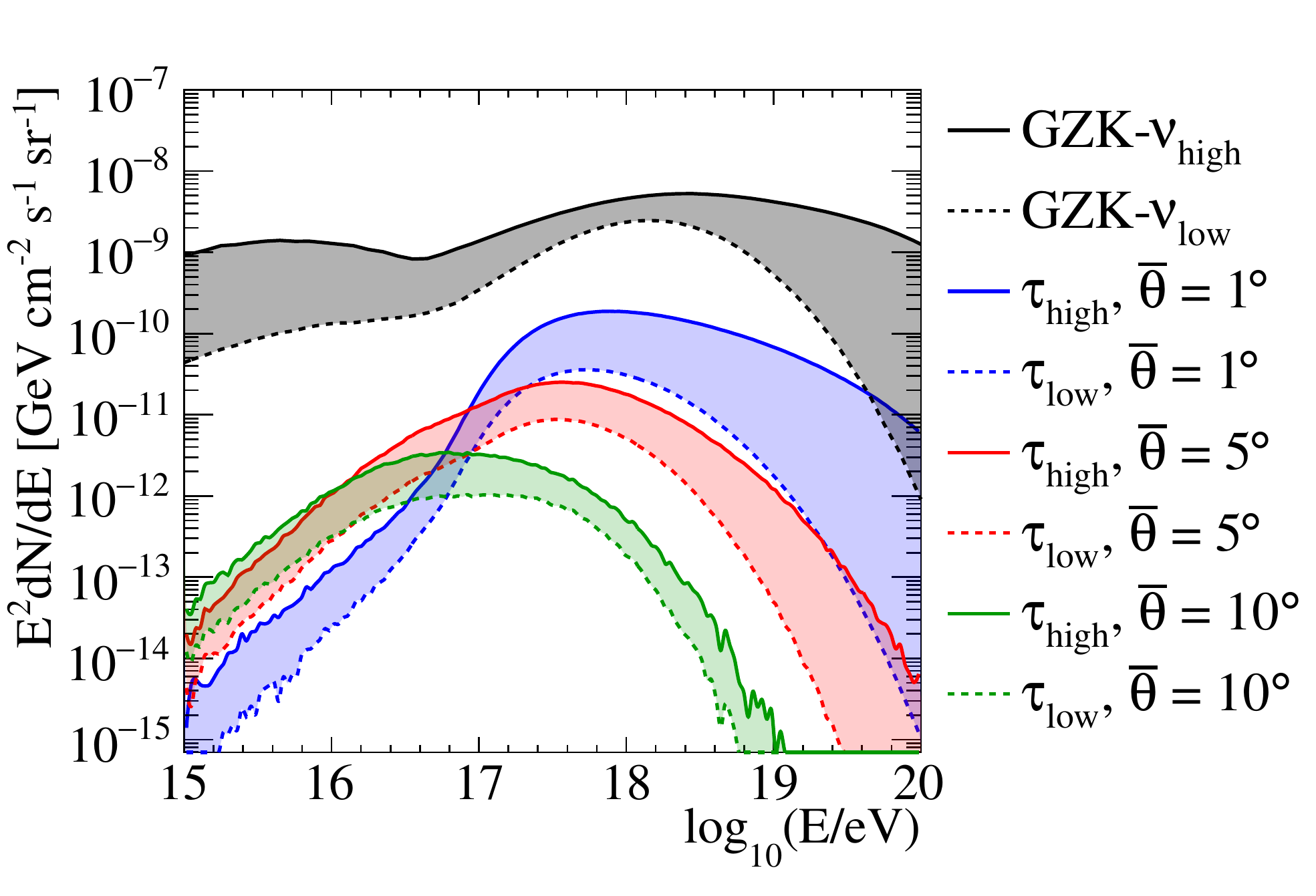} %
   \caption{The range of cosmogenic neutrino fluxes from Kotera~2010~\cite{Kotera_2010} and the resulting flux of $\tau$ leptons for emergence angles $\bar\theta=$ $1^\circ$, $5^\circ$, and $10^\circ$ (see Figure~\ref{fig:tau_geom}). The results use the middle neutrino-nucleon cross-section curve (Figure~\ref{fig:cross_sections}), ALLM energy loss rate (Figure~\ref{fig:tau_E_loss}) and $D=4$~km thick ice.}
   \label{fig:Kotera_tau_flux}
\end{figure}

So far we have treated the behavior of mono-energetic $\nu_\tau$'s. In this section, we characterize the exiting flux of $\tau$ leptons given a cosmogenic neutrino flux model. In Figure~\ref{fig:Kotera_tau_flux} we show the Kotera~2010~\cite{Kotera_2010} range of neutrino flux models along with the exiting $\tau$ lepton fluxes for emergence angles $\bar\theta=$ $1^\circ$, $5^\circ$, and $10^\circ$. We have used the middle neutrino-nucleon cross-section curve and the ALLM $\tau$-lepton energy loss rate model with a 4~km thick ice sheet.  
From Figure \ref{fig:Kotera_tau_flux} it is clear that 
$\tau$ leptons with the highest energy exit only at the smallest emergence angles. As the emergence
angle increases, these $\tau$ leptons are suppressed with
an enhancement of the $\tau$ lepton flux at 
$\sim 10^{17}$~eV and below. Note also that at the highest
energies $10^{19} - 10^{20}$~eV, the $\tau$ lepton decay length is
$490 - 4900$~km, respectively, meaning that a high altitude detector observing towards the horizon could potentially have a significantly improved sensitivity to these energies compared to a detector on the ground. Also note that at emergence angles of $10^{\circ}$, a ground-based detector is not likely to be able to observe a $\tau$-lepton decay since the observables (Cherenkov optical or radio emission and secondary particles) are forward directed. Fluorescence emission could potentially be observed at these geometries but these detectors typically have low duty cycles. 

\begin{figure}[!ht]
  \centering
   \includegraphics[width=1.0\linewidth]{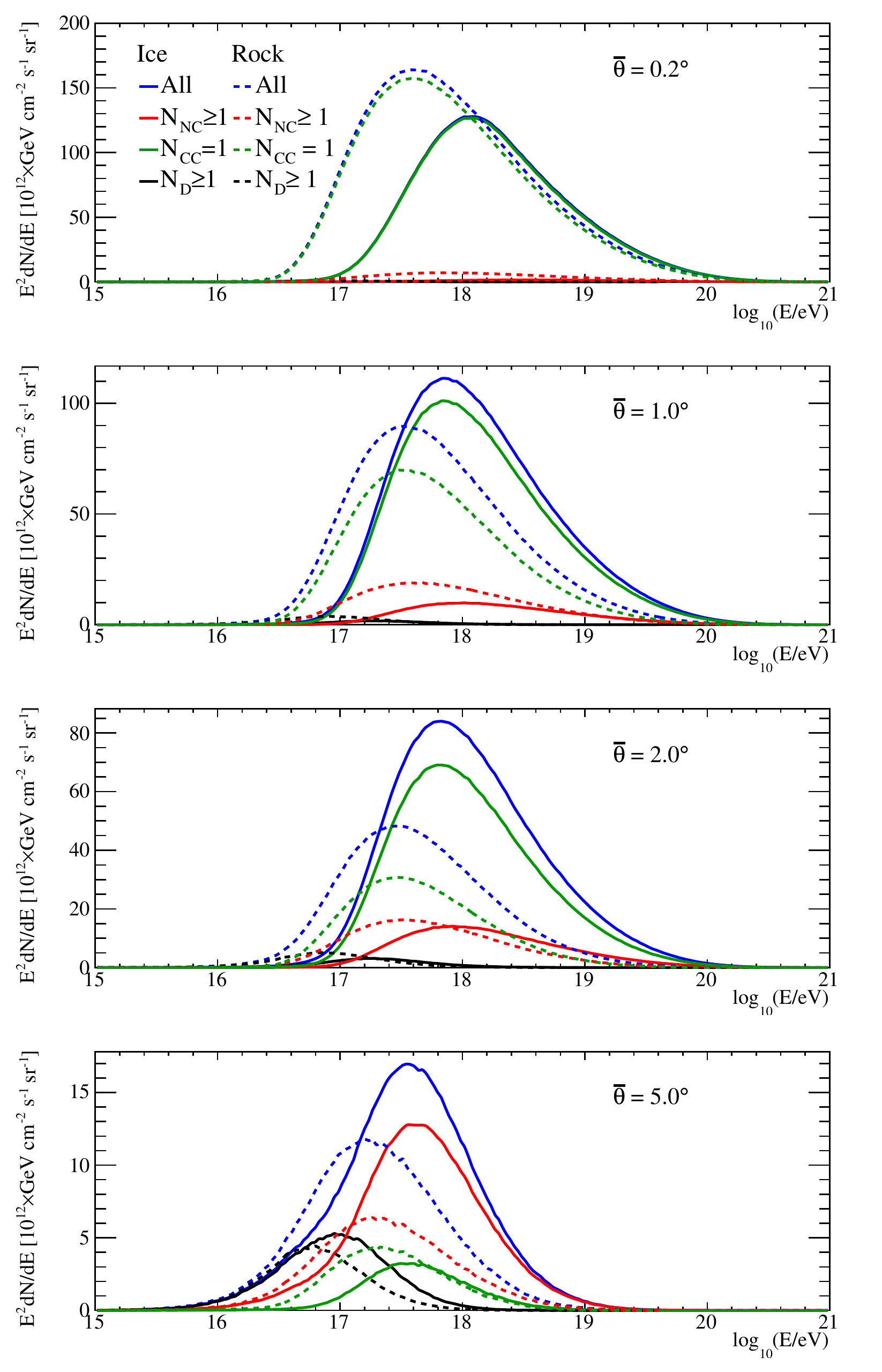} %
   \caption{The resulting flux of $\tau$ leptons for a cosmogenic neutrino flux in the middle of the flux ranges from Kotera~2010~\cite{Kotera_2010} (Grey band Figure \ref{fig:Kotera_tau_flux}). The different line colors indicate the interaction history that led to the exiting $\tau$ leptons (see text for more details). We show the effect of a 4~km thick ice layer (dashed lines) versus bare rock (solid lines) for 4 different emergence angles as indicated on the panels. These results are obtained using the middle neutrino-nucleon cross-section curve and ALLM energy loss rate.}
   \label{fig:Channels_RockIce}
\end{figure}

Finally we study how the different features contribute in concert for a given 
incident neutrino flux. 
For this purpose, we compare in Figure \ref{fig:Channels_RockIce} 
 propagation through the Earth with a 4 km thick ice-layer
(0.92~g~cm$^{-3}$) to propagation through regular rock 
(2.6~g~cm$^{-3}$) for a neutrino flux centered in the 
previously discussed band~\cite{Kotera_2010}.  
We display the total exiting $\tau$-lepton flux at several emergence angles and the fractions of the flux with a particular propagation history. As the number of $\nu_\tau$ interactions in the Earth increases, the propagation history of the exiting $\tau$ leptons becomes increasingly complex and therefore we grouped the exiting $\tau$ leptons in the three following categories:


\begin{description}
\item [N$_{\text{CC}}=1$:] exiting $\tau$ leptons resulting from a single $\nu_\tau$ charged current interaction (green lines). 
\item [N$_{\text{NC}}\geq1$:] exiting $\tau$ leptons produced by a $\nu_\tau$ that underwent at least one neutral current interaction (red lines). 
\item [N$_{\text{D}}\geq1$:] exiting $\tau$ leptons where at least one $\tau$ decay occurred in the chain of events leading to its production (black lines). 
\end{description}
The $N_{\text{NC}}\geq1$ and  N$_{\text{D}}\geq1$ fractions are not exclusive, but are chosen to illustrate when these two channels of $\nu_{\tau}$ regeneration become important. 

For very small emergence angles ($\bar\theta<$~0.25$^\circ$) $\nu_{\tau}$
regeneration is negligible as can been seen in top panel in Figure
\ref{fig:Channels_RockIce}. However 
at $\bar\theta \simeq 1^\circ$ the regeneration of
$\nu_{\tau}$'s becomes significant. In rock 20\,\% (35\,\%) of the exiting
$\tau$ leptons exhibit in their propagation history one or more neutral
current interactions at $\bar\theta = 1^\circ$ ($\bar\theta
=2^\circ$). Because the density in ice is lower than in rock, 
  the onset of regeneration in ice through neutral current interactions happens at
larger emergence angles. The fraction that underwent at least a single neutral current interaction in ice is 8\,\% at $\bar\theta = 1^\circ$, but increases rapidly to 30\,\% at $\bar\theta = 2^\circ$. 
The regeneration through decaying $\tau$ leptons becomes a significant
contribution only at larger emergence angles 
($\bar\theta = 2^\circ$) as can be observed from the lines labeled $N_D\geq1$ in the bottom 
panel of Figure \ref{fig:Channels_RockIce}.

Due to lower energy losses and lower interaction probabilities, the total exiting $\tau$ lepton flux peaks always at higher energies for the ice layer with respect to rock. Also, only at very small emergence angles  ($\bar\theta \lesssim 0.3^\circ$) the maximum energy flux ($E^2dN/dE$) of exiting $\tau$ leptons that propagated through rock exceeds that of a 4~km ice-layer. At these angles, the integrated density of the ice provides less charged current interactions than in rock. At larger angles, the ice layer is the more efficient medium to convert the neutrino flux into exiting $\tau$ leptons for neutrino energies $>3\times 10^{18}$~eV due to the lower $\tau$ lepton energy loss rate (see Section IV.C), giving a strong preference to experiments that would monitor ice or oceans for exiting $\tau$ leptons from ultra-high energy $\nu_\tau$'s at small emergence angles.   



\section{Discussion and Conclusions}
\label{sec:discussion_and_conclusions}
We have presented and characterized a $\nu_\tau$ and $\tau$ lepton propagation simulation that includes the effects of ocean or ice layer on Earth as well as regeneration and variations on models of neutrino-nucleon interaction and $\tau$ lepton energy loss rate. 
%
%
The results of these simulations show that for $\nu_\tau$ energies exceeding $3\times 10^{18}$~eV, the probability that a $\tau$ lepton exits the surface of the Earth into the atmosphere can be significantly boosted by the presence of a kilometer-scale thick layers of water or ice. 
The boost in the probability of exiting $\tau$ leptons is accompanied with a boost in energy and can be further enhanced depending on the neutrino-nucleon cross-section and $\tau$ lepton energy loss rate model. 
However, for $\nu_\tau$ energies below $3\times 10^{18}$~eV, we have found that bare rock has an advantage due to the $\tau$-lepton range being dominated by the decay lifetime, which is density-independent, rather than energy loss, which is density-dependent.
In this study, we have also considered uncertainties consistent with the standard model and found significant variations.

Given the magnitude and variation of the exiting $\tau$ fluxes due to interaction uncertainties, they should be considered in the reported results of various experiments. It is possible that observatories on the ground or at altitude could have significantly different sensitivities depending on the choice of model even within standard model uncertainties. In particular, ground based observatories are most sensitive to small emergence angles, where the dependence on neutrino-nucleon cross-section can be most pronounced. The potential benefit of observing at high altitude is to extend the range of emergence angles. Increasing the altitude of the detector enables observation of $\tau$ leptons exiting at larger emergence angles. 

Observing at altitude could also be of interest for constraining the
neutrino-nucleon cross-section and $\tau$ lepton energy loss rates. With a
sufficiently high neutrino event rate, the relative fraction of observations
and energies as a function of emergence angle in the presence of a layer of
water could be used to estimate the particle interaction parameters. An 
assessment of the sensitivity to these parameters will depend
principally on detector altitude and on the existence of a
water or ice layer, but ultimately on the detector characteristics as well as  on the neutrino flux, which is currently unknown at these energies. 

Consideration of the effects presented in this study should be of interest for
future detectors proposing to constrain the UHE neutrino flux using the $\tau$
lepton air shower channel. GRAND is proposing to place a large array of
antennas on mountainous terrain on the ground, which will
present some altitude variations. CHANT, on the other hand, is proposing to
observe these showers from stratospheric balloons or satellites, with the bulk
of their sensitivity over oceans. Radio detection experiments at
some altitude could also potentially observe $\tau$ leptons over various ice thicknesses and over the ocean.


\bigskip
\vspace{+10pt}
{\it Acknowledgements} Part of this work was carried out at the Jet Propulsion Laboratory, California Institute of Technology, under a contract with the National Aeronautics and Space Administration.
J. A-M and E.Z. thank Ministerio de Econom\'\i a (FPA 2015-70420-C2-1-R), 
Consolider-Ingenio 2010 CPAN Programme (CSD2007-00042),  
Xunta de Galicia (GRC2013-024 and ED431C 2017/07), Feder Funds,  
$7^{\rm th}$ Framework Program (PIRSES-2009-GA-246806) 
and RENATA Red Nacional Tem\'atica de Astropart\'\i culas (FPA2015-68783-REDT). W.C. thanks grant \#2015/15735-1, S\~ao Paulo Research Foundation (FAPESP).
We thank N. Armesto and G. Parente for fruitful discussions on the neutrino cross-section and $\tau$ lepton energy-loss models.
\bigskip


\end{document}